\documentclass[12pt]{article}
\usepackage{amsmath}
\usepackage{graphicx}
\usepackage{enumerate}
\usepackage{natbib}
\usepackage{url} 

\usepackage{amssymb, amsfonts, amsxtra}
\usepackage{epsfig, epsf, epic, fontenc} 
\usepackage{fancybox, lscape, subfigure} 
\usepackage{bm}
\usepackage{bbm}
\usepackage{hyperref}
\usepackage{xcolor}
\usepackage{physics}
\usepackage{multirow}
\usepackage{authblk}
\usepackage{dsfont}
\usepackage{float}
\usepackage{tikz}
\usepackage{pgfplots}
\pgfplotsset{compat = newest}
\usetikzlibrary{decorations.pathreplacing,calligraphy}

\newcommand{\asection}[2]{
\setcounter{section}{#1}
\addtocounter{section}{-1}
\section{#2}
}

\begin{document}

\def\spacingset#1{\renewcommand{\baselinestretch}%
{#1}\small\normalsize} \spacingset{1}

\begin{center}
{\LARGE\bf Robust inference for geographic regression discontinuity designs: assessing the impact of police precincts}
\end{center}

\begin{center}
Emmett B. Kendall$^1$, Brenden Beck$^2$, and Joseph Antonelli$^3$ 
\end{center}

\begin{center}
\small
$^1$Department of Statistics, North Carolina State University\\
$^2$School of Criminal Justice, Rutgers University Newark\\
$^3$Department of Statistics, University of Florida
\end{center}

\bigskip
\begin{abstract}
We study variation in policing outcomes attributable to differential policing practices in New York City (NYC) using geographic regression discontinuity designs (GeoRDDs). By focusing on small geographic windows near police precinct boundaries we can estimate local average treatment effects of police precinct practices on arrest rates. We propose estimands and develop estimators for the GeoRDD when the data come from a spatial point process. Standard GeoRDDs rely on continuity assumptions of the potential outcome surface or a local randomization assumption within a window around the boundary. These assumptions, however, can easily be violated in real applications. We develop a novel and robust approach to testing whether there are differences in policing outcomes that are caused by differences in police precinct policies across NYC. Importantly, this approach is applicable to standard regression discontinuity designs with both numeric and point process data. This approach is robust to violations of traditional assumptions made, and is valid under weaker assumptions. We use a unique form of resampling to provide a valid estimate of our test statistic's null distribution even under violations of standard assumptions. This procedure gives substantially different results in the analysis of NYC arrest rates than those that rely on standard assumptions.
\end{abstract}

\noindent%
\vfill

\newpage
\spacingset{1} 

\section{Introduction}

Policing varies across political boundaries, such as state or city borders. Such differences are expected, but we know very little about whether smaller, sub-municipal boundaries like police districts, precincts, and service areas also influence police outcomes \citep{klinger1997negotiating}. This lack of research persists despite police officers reporting that their behavior and perception is influenced by precinct boundaries \citep{hassell2007variation}. Police have wide discretion when choosing to make an arrest, so arrest rates could be particularly susceptible to spatial variation \citep{herbert1996policing}. Understanding whether precincts police differently has important implications for equity and policy. Variation in policing between cities is tolerated because it results, in part, from the electoral choices of residents. Variation within cities, however, generates a more troubling kind of inequality. Residents do not vote for their police precinct commander and they expect treatment equal to that of people in other neighborhoods. Policing variation generated by differences in police precinct policies or practices might also compound other forms of spatial inequity like residential segregation, racial bias, or high crime areas \citep{bell2020anti}. 

Another potential consequence of between-precinct variation in policing is diminished policy efficacy. Some recent police reform efforts have attempted to reduce the number of pedestrian stops and frisks, reduce use of deadly force, and improve police-community relations. Most such reforms are implemented at the city scale, but if significant variation exists between police precincts, such a one-size-fits-all approach might fail. Even place-based interventions like hot-spots policing target high-crime areas and ignore precinct boundaries. Research has long understood the salience of micro places in shaping crime, but less is known about how local characteristics shape policing. It is likely that police behavior, like criminal behavior, varies greatly by place. \textcolor{black}{In this study, we examine the 77 precincts of the New York City Police Department (NYPD) to determine if arrest rates differ across precincts, thus providing insight into potential arresting practice differences among officers in different precincts.}

This is a difficult question to answer because the regions each police precinct covers are different from one another with respect to important demographic and criminological variables. One precinct might have different low-level arrest rates than another because it has higher crime rates, more targets for theft, or more transient populations. Therefore, we have to isolate the effect of the police precinct practices or policies themselves. Randomization is the gold standard for drawing causal conclusions, but while these are occasionally available in the criminology literature to evaluate policies like hot spots policing \citep{puelz2019graph}, in many scenarios they are not available or feasible. When evaluating the impact of police precinct practices, we can not randomize individuals to a police precinct by forcing them to live or work in certain areas of a city. The ubiquity of observational studies has led to a wide range of approaches to estimate causal effects under as weak of assumptions as possible. Common approaches are difference-in-difference estimators \citep{ashenfelter1984using, lechner2011estimation}, the regression discontinuity design \citep{thistlethwaite1960regression,imbens2008regression, cattaneo2019regression}, interrupted time series analysis \citep{cook1979quasi, bernal2017interrupted}, and synthetic control analysis \citep{abadie2010synthetic}, among others. In the context of policing and criminology, these ideas have been used to address important issues, such as whether increased oversight of police leads to increases in crime and decreased effectiveness of the police force \citep{ba2019effect}, quantifying the impact of a penalty system for drivers in Italy on traffic incidents and traffic-related fatalities \citep{de2013deterrent}, or estimating the heterogeneous effects of neighborhood policing \citep{antonelli2020estimating, beck2020effects}.

In this study we focus on the regression discontinuity design and its extensions to geographic settings and point process data. For an in-depth review of standard regression discontinuity designs and implementation details, see \cite{imbens2008regression} and \cite{cattaneo2019regression}. The traditional regression discontinuity design occurs when treatment assignment is either partially or completely determined by a pre-treatment covariate, typically referred to as the running or score variable. There exists a cutoff value of this running variable, above which units receive treatment, and below which units receive the control. The fundamental idea is that units within a small distance around the cutoff value form a locally randomized experiment \citep{mattei2017regression}. The estimand of interest is a local treatment effect at the cutoff value, and nearby observations are used to extrapolate what would happen both under treatment and control at this boundary value. This approach has been extended to multivariate running variables such as the results of two types of educational tests \citep{matsudaira2008mandatory}. A specific example of a bivariate running variable is found in the GeoRDD where latitude and longitude are used to determine treatment assignment. Important aspects specific to the geographic design are highlighted in \cite{keele2015geographic}. This design has been used to estimate the effect of private police departments on crime \citep{macdonald2016effect}, the impact of voter initiatives on voter turnout \citep{keele2015enhancing}, the effect of the Civil Rights Act of 1875 \citep{harvey2020applying}, and whether school districts impact housing prices \citep{rischard2020school}. 

Regression discontinuity designs rely on assumptions that the potential outcomes are smooth at the cutoff value or that treatment behaves as if it were randomized within a window around the cutoff value. To assess the validity of these assumptions, a number of falsification tests have been proposed. A negative control approach is to treat an observed covariate as an outcome, where we know the treatment should not affect this covariate, and estimate the treatment effect on this covariate to see if the approach correctly estimates a null association \citep{lee2004voters}. Another issue is that the running variable can be manipulated by subjects if they are aware of the cutoff value, and this can be evaluated by checking if the distribution of the running variable is continuous at the cutoff \citep{mccrary2008manipulation, cattaneo2017comparing}. Other approaches examine the sensitivity of results to bandwidth selection \citep{lemieux2008incentive}, as robustness of results to this choice provides increased belief in the resulting findings. 

\textcolor{black}{ In this work, we develop two key contributions to the literature on geographic regression discontinuity designs. For one, we extend the GeoRDD literature to settings where the outcome is a point process, and causal estimands and assumptions are defined in terms of intensity surfaces. Standard GeoRDD approaches do not apply here because they are designed for settings with a numeric outcome (such as voting behavior or test scores) measured at spatial locations. In contrast, we do not have a numeric outcome; instead, our outcome is the location of events, and we are interested in studying the expected number of events in specific subsets of the spatial domain. This necessitates a modification of key assumptions and estimands, and relies on distinct estimation strategies that we develop. Our second contribution, which is applicable to general GeoRDD settings, not just the point process one seen here, is that we provide valid hypothesis tests for causal estimands under certain violations of existing assumptions typically utilized in the GeoRDD. By using a novel resampling scheme, our approach allows for violations of the assumptions that treatment is as if randomized within a window around the cutoff point of the running variable, or that the potential outcomes are smooth at the cutoff. We exploit a large spatio-temporal data set of crime and arrest data in NYC to find streets that behave similarly to precinct boundaries, but by definition have no effect of police department practices as they are fully contained within a single police precinct. We use these streets to construct a null distribution that accounts for violations of local randomization or continuity assumptions and provides a valid hypothesis test of individual precinct effects, as well as a test for the overall degree of variation in policing across NYC. }

\textcolor{black}{The relevant \texttt{R} code and data for reproducing all numerical results presented in this article are available at \href{https://ebkendall.github.io/research.html}{\it ebkendall.github.io/research.html}.}


\section{Policing data in NYC and preliminary analyses}
\label{sec:data}
Our analyses draw on two data sources made public  by the NYPD: NYPD Arrest Data and NYPD Complaint Data. Both provide information at the incident level with geolocated, address data for all arrests and crimes reported to the police in the years 2010-2018. The NYPD is divided into 77 police precincts, each patrolling a particular geographic area of the city. We exclude the precinct corresponding to Central Park, which does not have any residents, leading to 76 precincts in our analysis.  Our goal is to use these data to understand whether there is variability in arresting practices across police precincts in NYC, and whether individuals are more or less likely to be arrested depending on which precinct's police force they are exposed to. \textcolor{black}{Before describing our problem in more detail, it is important to emphasize that police precincts can refer to both geographic areas, as well as the police organizational unit that polices that geographic area. Whenever we refer to effects of police precincts, we are referring to effects of decisions, policies, or practices of the police department within that precinct, not the effects of the geographic area itself. This is a well-defined treatment variable of interest to study, because police departments have different police commanders, different policies, and other features that may affect arrest rates. }

\begin{figure}[!htb]
    \spacingset{1}
    \centering
    \includegraphics[width = 0.45\linewidth]{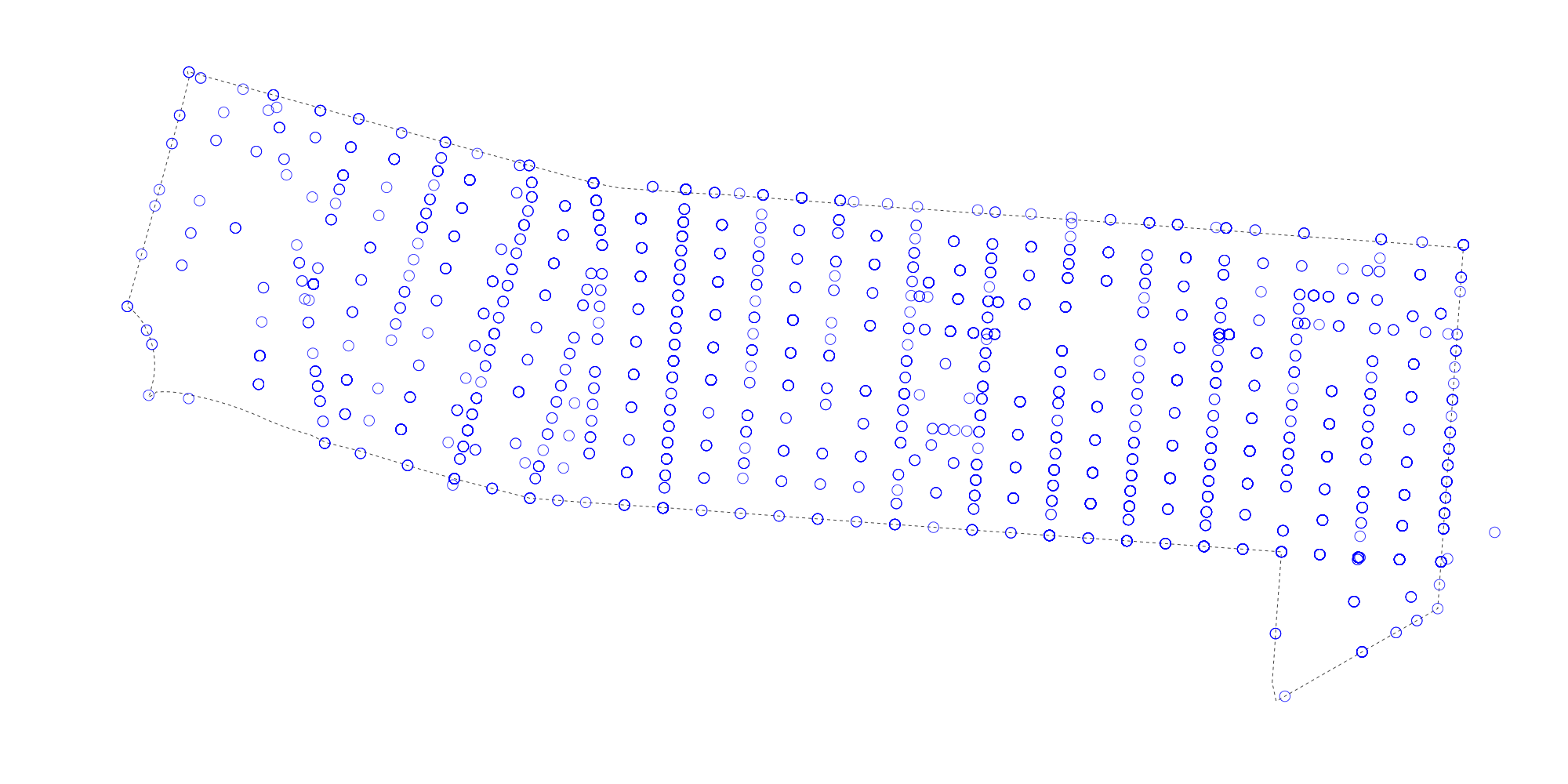} \includegraphics[width = 0.45\linewidth]{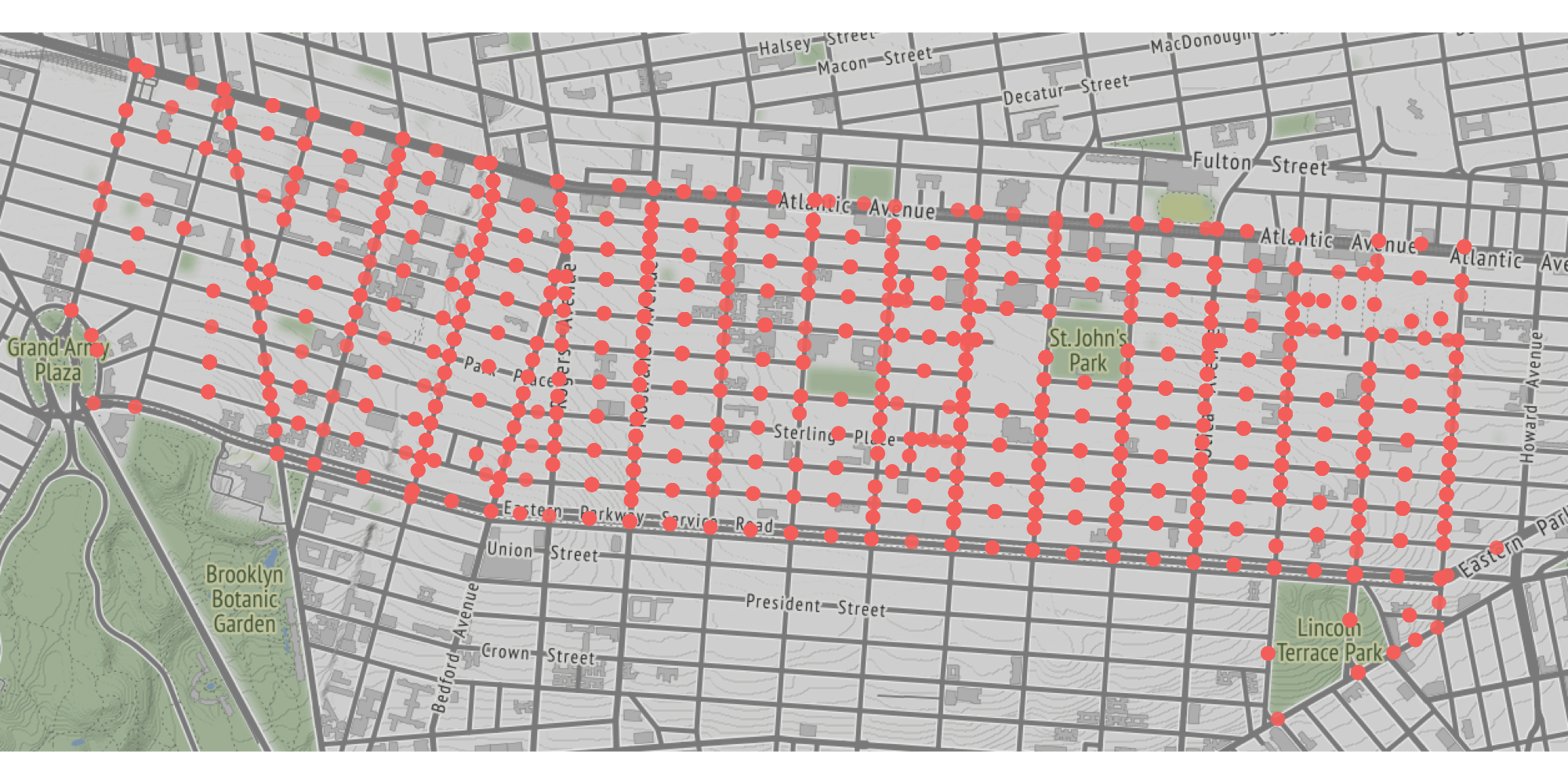}
    \caption{Arrest locations from officers of Precinct 77 in NYC during the year 2014.}
    \label{fig:Precinct77}
\end{figure}

Using these data, we can visualize both when and where arrests occur as well as the precinct from which the arresting officer originates. Figure \ref{fig:Precinct77} highlights the arrest data for Precinct 77 in NYC during the year 2014, both with and without the roadmap of the city overlaid on the figure using the \texttt{R} package \texttt{ggmap}  \citep{ggmap2013}. 
This figure reveals features of the data we will leverage in our approach to inference. Both arrests and crimes fall directly on streets, and are not spread out across the entire spatial domain. \textcolor{black}{Note that because the observed spatial locations are restricted to a grid, a single spatial location can correspond to multiple arrests or crimes.} Additionally, precinct boundaries are streets themselves, and there are arrests that fall directly on these boundaries. 

\begin{figure}[!htb]
    \spacingset{1}
    \centering
    \includegraphics[width = 0.45\linewidth]{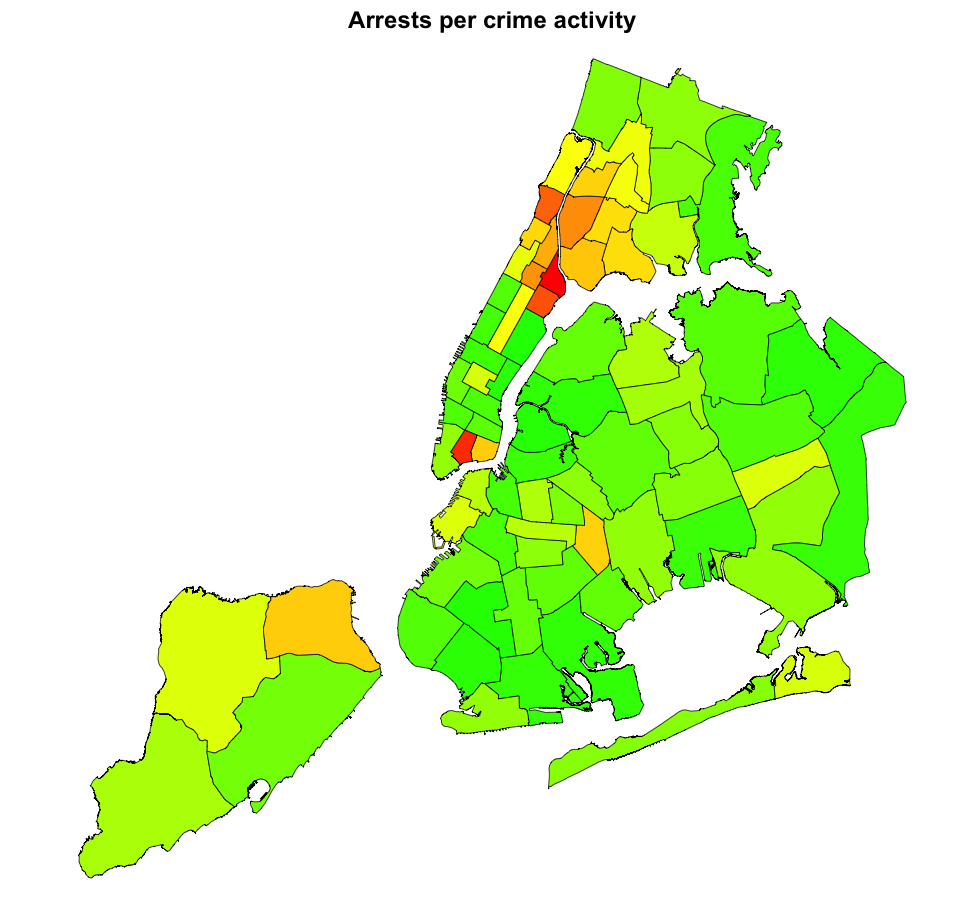} 
    \caption{Number of arrests divided by the number of crimes over the entire study period, separated by precinct}
    \label{fig:ArrestRatesPrecinct}
\end{figure}

Suppose that we are interested in estimating whether there is a difference in arresting practices between two neighboring police precincts. Figure \ref{fig:ArrestRatesPrecinct} plots the number of arrests divided by the number of crimes within each precinct, and we can see significant variation across police precincts. This could be driven by many factors, such as the types of crime committed within each precinct. Alternatively, it could be that separate police precincts have distinct policies or practices that lead to such variation, and our goal is to study whether it is the latter. Formally, we want to assess whether there is a causal effect of police precinct on arrest rates. Given the lack of rich covariate information, but high degree of spatial resolution, one might choose to use a GeoRDD to estimate the effect of police precincts between two neighboring regions. The GeoRDD leverages the fact that nearby areas should be similar with respect to important unobserved characteristics that are associated with arrest rates. In this context, this assumption would be satisfied if areas on either side of the border between the two precincts are similar to each other. \textcolor{black}{If individuals generally do not choose where to live based on which police precinct their home falls in, which is a reasonable assumption for many police precinct boundaries, then differences in arrest rates would be attributable to differences in police precinct practices.}

\begin{figure}[!htb]
    \spacingset{1}
    \centering
    \includegraphics[scale=0.17,trim={0cm 5cm 35cm 6cm},clip]{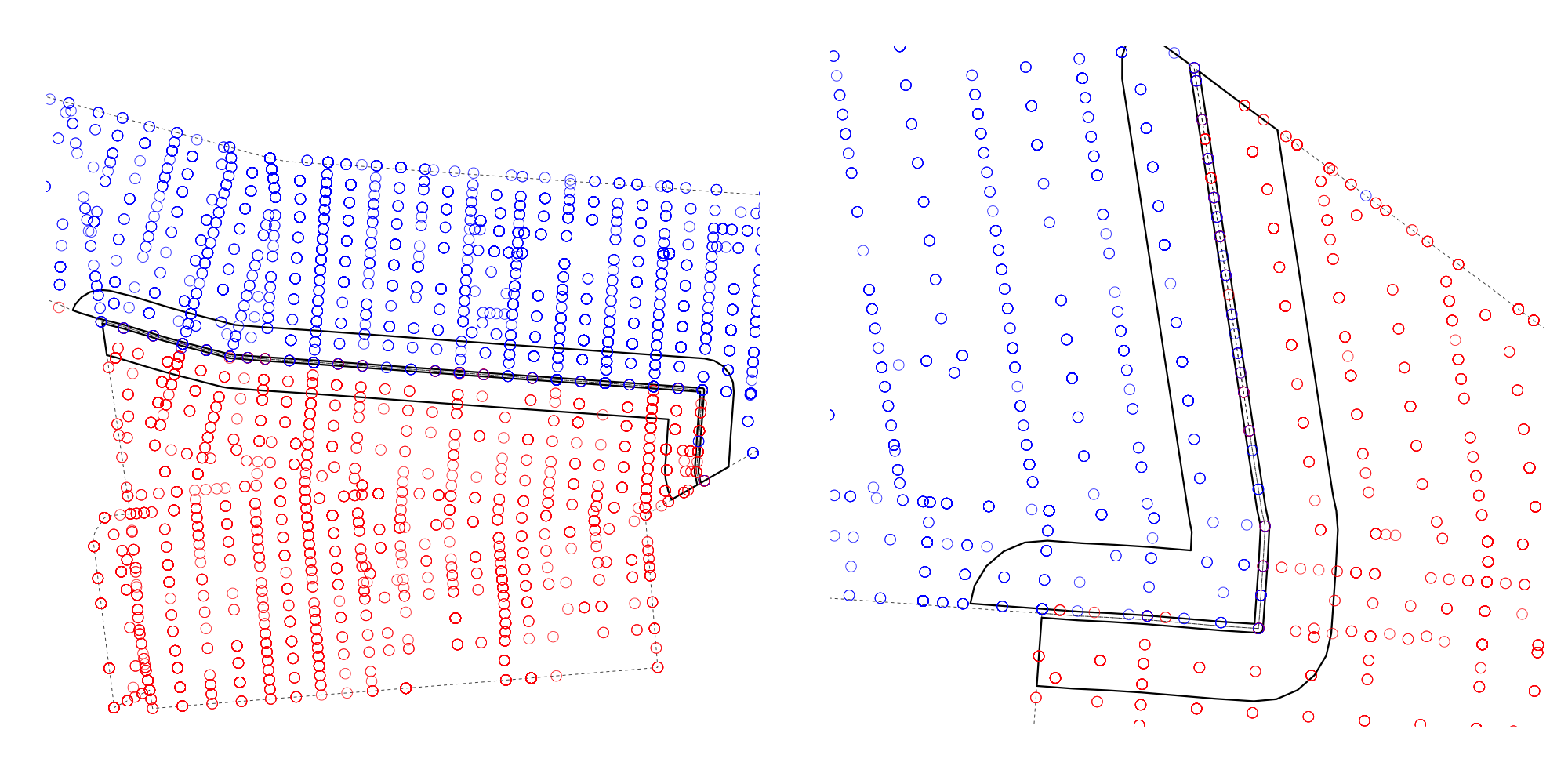}
    \caption{Arrest locations near the border of Precincts 71 and 77.}
    \label{fig:Precinct77_79}
\end{figure}

Existing GeoRDD methodology does not immediately apply to the point process data seen here. We formalize estimands and methodology for estimation in this setting in Section \ref{sec:MethodologyOverall}, but as a preliminary analysis we can draw a buffer around the border of two precincts and study the difference in arrest rates made by each precinct within the buffer over time. An illustration of the setup can be found in Figure \ref{fig:Precinct77_79}, where arrests are color coded based on the arresting officer's precinct. If there is no variation in policing practices across precincts, then under certain causal assumptions detailed in Section \ref{sec:MethodologyOverall}, we would expect to find a significant difference in arresting practices between two neighboring precincts with probability $\alpha$, where $\alpha$ is the pre-specified type I error. Given that there are 144 precinct-precinct borders of interest, we would expect to see roughly $\alpha \times 144$ significant associations and the distribution of p-values across these tests to be roughly uniformly distributed if police precincts do not affect arrest rates.

\begin{figure}[!htb]
    \spacingset{1}
    \centering
    \includegraphics[width=0.49\linewidth]{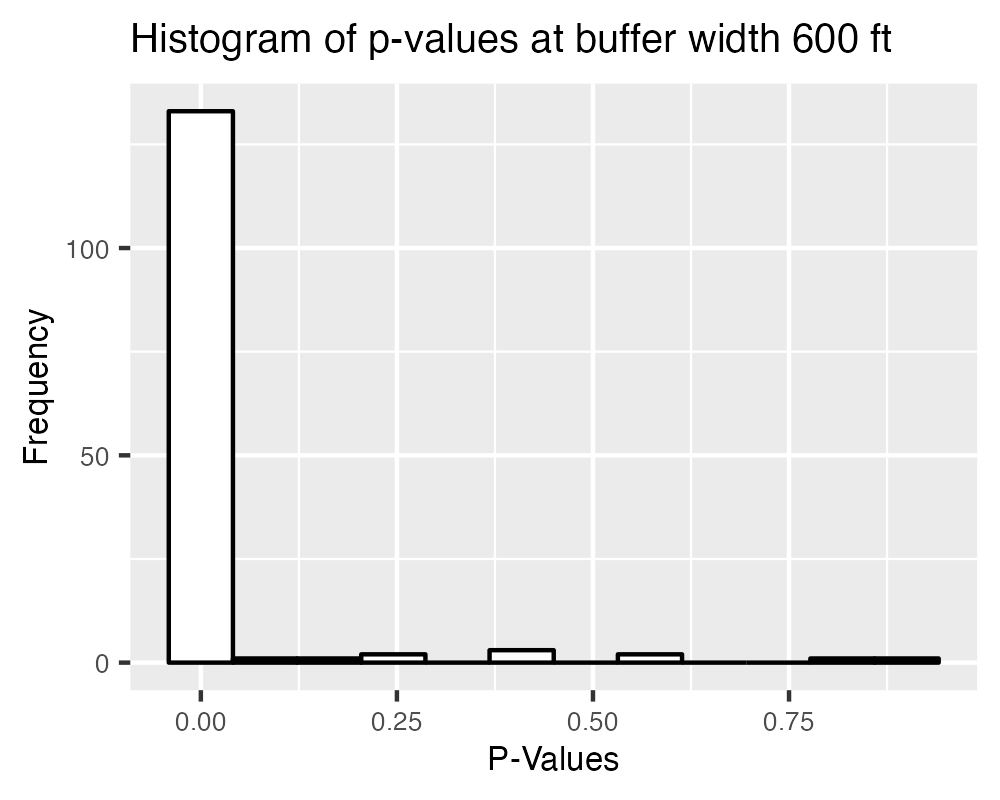}
    \includegraphics[width=0.49\linewidth]{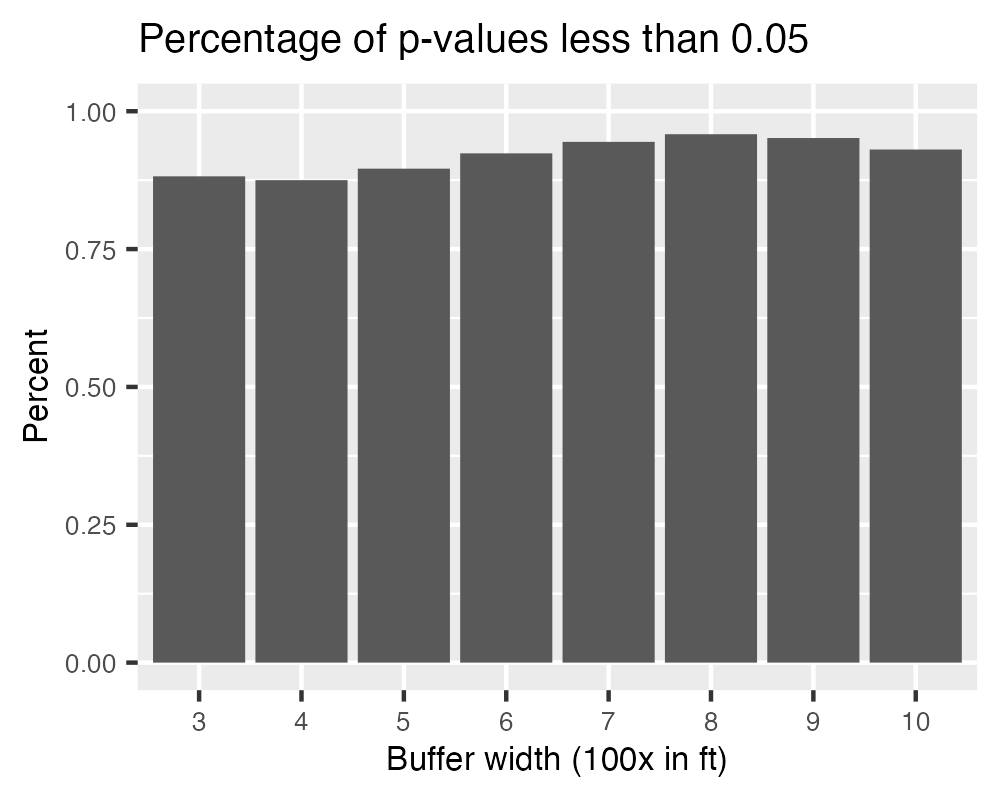}
    \vspace{-.1in}    
    \caption{Distribution of p-values across all 144 borders at a buffer width of 600 feet (left) along with the percent of rejections as a function of buffer size around the border (right) for counts of arrests.}
    \label{fig:NaivePvalue}
    \vspace{-0.1in}
\end{figure}

Suppose buffers with radius lengths of 300 to 1000 feet are drawn around each border. Then, the number of arrests made by each of the two precincts within the bounds of the buffer zone are counted so that we can test if there is a significant difference in arrest counts. Specifically, if we let $(Y_0, Y_1)$ be the number of arrests on the two sides of the border, we expect that $Y_0 \sim \text{binomial}(Y_0 + Y_1, 0.5)$ under the null hypothesis of no precinct effect.  We can use this result to test the hypothesis that the expected number of arrests on either side of the boundary is the same. Figure \ref{fig:NaivePvalue} shows the distribution of p-values from this test for a 600 foot buffer, as well as the percentage of rejections out of 144 borders as a function of the buffer width. We see that 92.4\% of the p-values are less than 0.05 when we use a buffer radius of 600 feet, with similar percentages for other buffer widths, and the p-value histogram is far from uniform. This may occur because: (1) there truly are large differences in arresting practices across NYC police precincts, (2) either the causal assumptions do not hold or the statistical test is invalid, or (3) some combination of these two.

\subsection{Negative control analyses}
\label{sec:DataNegControl}
\begin{figure}[!htb]
    \spacingset{1}
    \centering
    \includegraphics[width=0.49\linewidth]{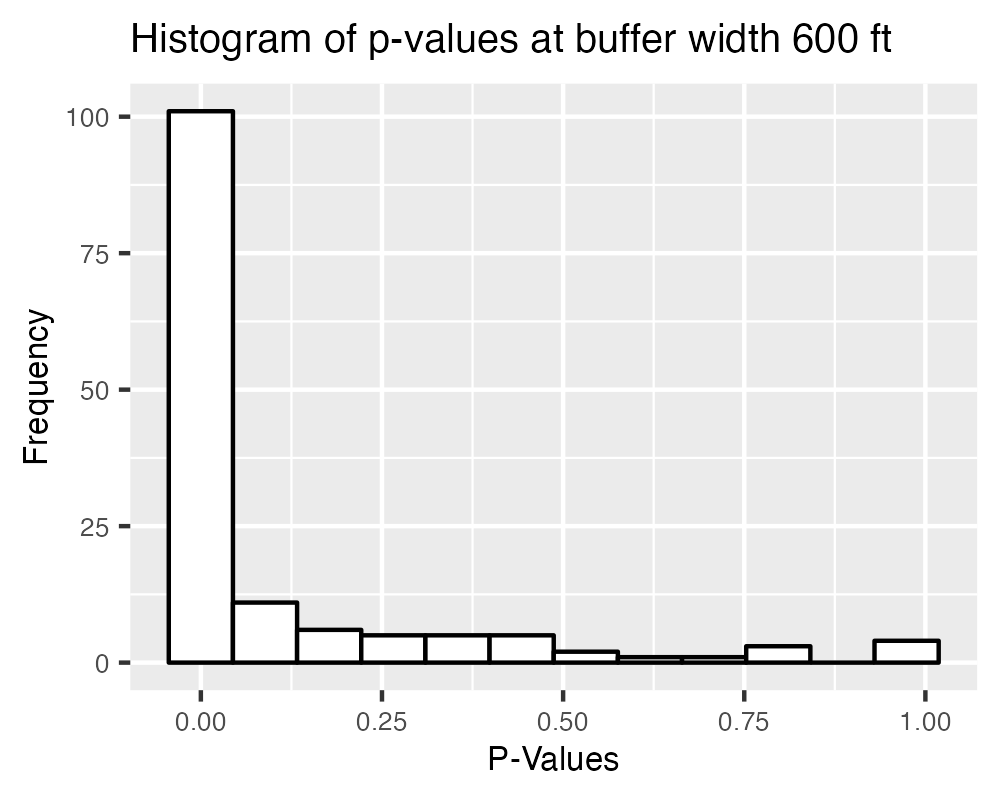}
    \includegraphics[width=0.49\linewidth]{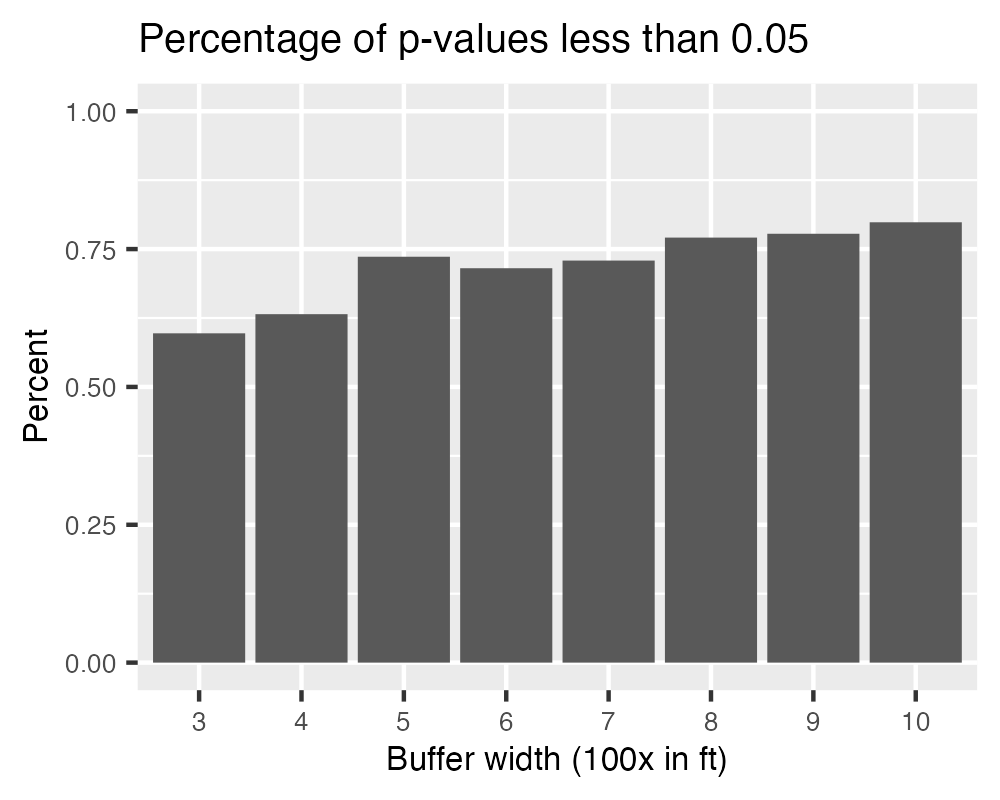}
    \vspace{-.1in}
    \caption{Distribution of p-values across all 144 borders at a buffer width of 600 feet (left) along with the percent of rejections as a function of buffer size around the border (right) for the negative control outcome looking at counts of trees.}
    \label{fig:negControlNaive}
    \vspace{-0.1in}
\end{figure}

While we cannot know whether the significant effects on arrest rates are due to a true causal effect of police precincts or violations of key assumptions, we can examine similar data that \textcolor{black}{is known to have no police precinct effect}. As a negative control outcome, we consider the 2015 Street Tree Census data set provided by \textit{NYC Open Data}. This data set provides the location of all trees in NYC, and has a similar structure as the crime/arrest data as all locations are recorded at street locations. Clearly police departments do not decide where to place trees in a precinct, which might lead one to assume that the difference in tree counts on either side of any two bordering precincts is expected to be small. This intuition is not correct in this instance, as Figure \ref{fig:negControlNaive} shows very similar results to that of the arrest data. The p-value distribution across precinct boundaries is highly skewed, and the percentage of significant findings, even for small buffer widths of 300 feet, is well above the desired type I error rate.
This motivates the need to develop a methodology that is robust to violations of assumptions on the similarity of the populations on both sides of precinct boundaries, and that is able to provide valid inference in this setting.

\section{Causal inference for spatial point processes with GeoRDDs}
\label{sec:MethodologyOverall}

The observed data on our outcome of interest consists of a set of geographic locations corresponding to each arrest in NYC between 2010 and 2018. Our data can be thought of as being generated from a point process, in that an arrest will randomly occur at some time point, and at a given location. Since we are interested in studying variability in arrest rates across space, rather than time, we aggregate across the temporal component to focus solely on space. We denote geographical coordinates by $\boldsymbol{S} = (S_{1}, S_{2})$, which correspond to latitude and longitude.  Therefore the observed outcomes are given by $\mathcal{S} = \{\boldsymbol{S}_{1}, \boldsymbol{S}_{2}, \dots, \boldsymbol{S}_{N}\}$ where $\boldsymbol{S}_{1}, \boldsymbol{S}_{2}, \dots, \boldsymbol{S}_{N}$ represent each location at which an outcome is observed (e.g., where each arrest occurs). We use $|\mathcal{S}| = N$ to denote the cardinality of this set, which corresponds to the number of events in the entire spatial domain studied. \textcolor{black}{To reiterate, standard GeoRDD estimands and inferential strategies do not apply here because we do not observe a numeric outcome value at these locations. Instead, the outcome is the location of arrests in NYC, and its distribution is governed by an unknown point process.}

At times we focus on specific subregions of the entire spatial domain and therefore we can define corresponding region-specific quantities. Denote $R$ as a subregion under study, such as the region within 300 feet of the boundary between two precincts. Let $Y(R) = \sum_{i=1}^N \mathbf{1}(\boldsymbol{S}_{i} \in R)$ represent the number of outcomes that occurred in region $R$. We assume that our data follow an inhomogeneous point process \citep{daley2003introduction} with intensity function given by $\lambda(\boldsymbol{s})$, where $\boldsymbol{s}$ denotes a spatial location. Specifically, this implies that 
\begin{equation}
\label{eq:exp_val}
    E(Y(R)) = \int_R \lambda(\boldsymbol{s}) d \boldsymbol{s},
\end{equation}
for any region $R$ in the domain of interest. 

\subsection{Single border estimands}
\label{subsec:sbe}
Now that we have introduced notation relevant to our problem, we can formally define potential outcomes and causal estimands in the context of spatial point process data. Our interest is in \textcolor{black}{whether arrest rates differ across police precincts throughout NYC}, and we can first answer this question by focusing on two adjacent precincts at a time, which we refer to as \textit{precinct 0} and \textit{precinct 1}. \textcolor{black}{Note that most precincts have more than one neighboring precinct; therefore, in the event that a precinct has, say, four neighbors, then there exists four distinct estimands for each adjacent precinct pair.} We can then extend these ideas to all adjacent precinct pairs in NYC in Section \ref{sec:Global}. First, let $\mathcal{B}$ denote the spatial boundary separating precinct 0 and precinct 1 (i.e., the border between the two precincts). Further, define a distance function $d(\boldsymbol{s}, \mathcal{B})$, which is the shortest distance between a point in space, $\boldsymbol{s}$, and the boundary, $\mathcal{B}$. Next, define the treatment variable $T(\boldsymbol{s})$ to be an indicator of whether a location is policed by police precinct 0 or 1. Clearly, this is a deterministic function of $\boldsymbol{s}$ as $T(\boldsymbol{s}) = \mathbf{1}(\boldsymbol{s} \in \text{precinct 1})$. This is referred to as a sharp regression discontinuity design as the forcing variable $(\boldsymbol{s}$ in this setting) completely determines the treatment assignment \citep{trochim1990regression}. With both the distance metric and treatment defined, we can more formally characterize the regions relevant to our study. Define
\begin{align*}
    R_{\delta, 0} &= \{ \boldsymbol{s}:d(\boldsymbol{s}, \mathcal{B}) < \delta, T(\boldsymbol{s}) = 0 \},\\
    R_{\delta, 1} &= \{ \boldsymbol{s}:d(\boldsymbol{s}, \mathcal{B}) < \delta, T(\boldsymbol{s}) = 1 \}.
\end{align*}
Intuitively, $R_{\delta, 0}$ is the region within distance $\delta$ of the boundary, $\mathcal{B}$, on the side of precinct 0, with an analogous interpretation for $R_{\delta, 1}$ and precinct 1. Then, denote $R_\delta$ to simply be the combined area of $R_{\delta, 0}$ and $R_{\delta,1}$ (i.e., the total region within distance $\delta$ of boundary $\mathcal{B}$). Figure \ref{fig:border_ill} provides an illustration of these terms.

\begin{figure}[!htb]
    \centering
    \begin{tikzpicture}[node distance={18mm}, thick, baseline={-.2cm}, scale = 2] 
    \coordinate (origin1) at (0,0);
    \coordinate (A1) at (-1,1);
    \coordinate (B1) at (4,1);
    \coordinate (C1) at (4,0.5);
    \coordinate (origin2) at (3,0);
    \draw (0,0) -- (-1,1);
    \draw (0,0) -- (-1,-0.5);
    \draw (3,0) -- (4,-1);
    \draw (3,0) -- (4, 0.5);
    \path [fill=cyan, opacity=0.6] (origin1) to  (A1) to (B1) to (C1) to (origin2);
    \path [fill=green, opacity=0.6] (origin1) to  (-1,-0.5) to (-1,-1) to (4,-1) to (origin2);
    \draw [dashed,ultra thick] (0,0) -- (3,0);
    \path[fill = red, opacity = 0.3] (0,-0.3) rectangle (3,0.3);
    \begin{scope}
        \clip (-1,-0.3) rectangle (0,0.3);
        \path[fill = red, opacity = 0.3] (0,0) circle(0.3);
    \end{scope}
    \begin{scope}
        \clip (3,-0.3) rectangle (4,0.3);
        \path[fill = red, opacity = 0.3] (3,0) circle(0.3);
    \end{scope}
    \node[draw] at (-0.3,-0.8) {precinct 1};
    \node[draw] at (3.3,0.8) {precinct 0};
    \node[above right] (1) at (-0.7,0) {$\mathcal{B}$};
    \draw[->] (1) to [out=340, looseness=1] (-0.05, 0); 
    \node[above right] (2) at (1.2,-0.02) {$R_{\delta, 0}$};
    \node[below right] (2) at (1.2, 0.01) {$R_{\delta, 1}$};
    \draw [pen colour={black}, decorate, 
    decoration = {calligraphic brace,
        raise=5pt,
        amplitude=5pt,
        aspect=0.5}, ultra thick] (3.2,0.3) -- (3.2,-0.3)
    node[pos=0.5,right=8pt,black]{$R_\delta$};
    \end{tikzpicture} 
    \caption{Illustration of the components surrounding the border.}
    \label{fig:border_ill}
    \vspace{-0.1in}
\end{figure}

We frame our problem and the regression discontinuity design within the potential outcomes framework \citep{rubin1974estimating}. We define $\mathcal{S}^1$ to be the set of locations with an arrest had every area been exposed to policing by police precinct 1 and $\mathcal{S}^0$ be the corresponding quantity for precinct 0. Accordingly, we can define $Y^1(R_{\delta})$ to be the number of outcomes we would observe in region $R_{\delta}$ if exposed to policing by precinct 1 and $Y^0(R_\delta)$ be the same quantity for precinct 0. We assume that these potential outcome point patterns come from inhomogeneous point processes with intensity functions $\lambda^1(\boldsymbol{s})$ and $\lambda^0(\boldsymbol{s})$, respectively. Precincts 0 and 1 could have different observed arrest rates for a number of reasons, many of which are \textit{not} due to a causal effect of police precincts. One area could have higher crime rates, different types of crimes committed with different clearance rates, or a different demographic of individuals in the population that live there. However, since precinct boundaries are defined on streets and many individuals are unaware of which precinct they reside in, we can focus our analyses on regions close to the boundary between the two precincts (i.e., $R_\delta$ for sufficiently small $\delta$) because the individuals on either side of the street are more likely to be similar and neighborhood characteristics should be more comparable. In particular, we examine a local average treatment effect defined by 
\begin{align}
  \theta(R_{\delta}) = E(Y^1(R_{\delta}) - Y^0(R_{\delta})) = \int_{R_{\delta}} \{ \lambda^1(\boldsymbol{s}) - \lambda^0(\boldsymbol{s}) \} d \boldsymbol{s}.  \label{eqn:LATE}
\end{align}
This treatment effect is local in the sense that it is the effect of being exposed to policing practices of precinct 1 versus precinct 0, but only in the region near the boundary, defined by $R_{\delta}$. This is standard in the regression discontinuity literature where treatment effects are typically identified at, or near, the cutoff of the forcing variable. 

Alternative estimands that provide more detailed information about the nature of the treatment effect are also of interest. One such estimand, which acknowledges that the treatment effect may vary spatially across the boundary of interest, can be defined as
\begin{align}
    \tau(\boldsymbol{b}) = \lambda^1(\boldsymbol{b}) - \lambda^0(\boldsymbol{b}) \quad \forall \ \boldsymbol{b} \in \mathcal{B}. \label{eqn:SpatialLATE}
\end{align}
This represents the difference in the underlying point process intensity surfaces at all locations on the boundary of interest, and is an extension of estimands seen in prior spatial regression discontinuity designs \citep{keele2015geographic} to the point process setting. This estimand allows for heterogeneity of the treatment effect across the boundary of interest. We can also aggregate this effect across the boundary using similar ideas as in \cite{keele2015geographic} and \cite{rischard2020school} by defining the following:
\begin{align}
    \tau = \frac{\int_{\boldsymbol{s} \in \mathcal{B}} w(\boldsymbol{s})\tau(\boldsymbol{s}) d \boldsymbol{s}}{\int_{\boldsymbol{s} \in \mathcal{B}} w(\boldsymbol{s}) d \boldsymbol{s}}, \label{eqn:WLATE}
\end{align}
where $w(\boldsymbol{s})$ is a weight function that assigns weight to each point on the boundary. Throughout, we assign equal weights, $w(\boldsymbol{b}) = 1$ for all $\boldsymbol{b}$, but the ideas used hold for any choice of weights. Other weights such as those that assign weight proportional to population size, or weights that minimize the variance of the estimated treatment effect, may also be of interest, though we refer readers to \cite{rischard2020school} for a broader discussion around this choice. Similar to (\ref{eqn:LATE}) this provides a local average treatment effect at the boundary, but we will see in subsequent sections that the identification assumptions and estimation strategies are slightly different between the two estimands. In general, we recommend using the point process estimand in equation \eqref{eqn:SpatialLATE} if one is interested in spatial heterogeneity of the effect across the border, and the average estimand in \eqref{eqn:WLATE} otherwise. In subsequent sections, we will see reasons for choosing these estimands over $\theta(R_{\delta})$. In particular, the identification assumptions are arguably more plausible and easier to justify (see Section \ref{sec:Identification}), and the selection of tuning parameters is more straightforward for estimating these estimands (see Section \ref{sec:SmoothingParameters}). Despite this, we still present results and discussion around the estimand in \eqref{eqn:LATE} as it is easier to study and provide intuition for mathematically, in addition to being computationally less demanding to estimate as it does not require estimating intensity surfaces or implementing cross-validation.

\subsection{Identifying assumptions}
\label{sec:Identification}

The main idea behind the regression discontinuity design is that by looking in a close window around the boundary, the two regions on either side of the boundary are very similar with respect to all important features except for which precinct's police department they are being policed by. \textcolor{black}{The impact of important confounding variables associated with arrest rates that may differ between the two precincts should be mitigated when looking within small geographic areas, as long as the confounding factors are continuous at the boundary between the two precincts.} We can therefore compare outcomes on either side of the boundary and attribute differences to the effect of the police departments in each precinct. Here we formalize this notion by explicitly writing down the assumptions by which the regression discontinuity design is able to identify the aforementioned local average treatment effects from the observed data. \textcolor{black} {First, for ease of exposition, we show these assumptions in the absence of additional covariate information. We discuss the role of covariates in detail in Section \ref{sec:Covariates}, and weaken the following identification assumptions to incorporate covariates in Supplementary Materials Section 2.} Additionally, in what follows we aim to weaken these assumptions and provide valid hypothesis tests even when some of these assumptions are violated. 

The two most commonly used assumptions unique to the regression discontinuity design are local randomization assumptions or assumptions on continuity of potential outcomes at the cutoff of the running variable. Local randomization states that treatment assignment is independent of the potential outcomes when looking only within a small window around the cutoff of the running variable \citep{mattei2017regression}. Continuity of potential outcomes is the most commonly used assumption in the regression discontinuity literature and specifies that the conditional mean of the potential outcomes under both treatment and control are continuous functions in the running variable \textcolor{black}{\citep{imbens2008regression}}. Here, we extend these assumptions to the spatial point process setting and show how they can be used to identify $\theta(R_{\delta})$ and $\tau(\boldsymbol{b})$, respectively.  First, we detail the assumptions needed to identify $\theta(R_{\delta})$, which are given in assumptions 1 and 2.

\noindent\textbf{Assumption 1:} \textit{Consistency of potential outcomes}. $$Y^t(R_{\delta, t}) = Y(R_{\delta, t}), \quad \text{for } t=0,1$$
\noindent\textbf{Assumption 2a:} \textit{Constant integrated intensity functions}. $$E(Y^t(R_{\delta, 1})) = E(Y^t(R_{\delta, 0})), \quad \text{for } t=0,1$$
Assumption 1 is a standard assumption required to link our observed data to the potential outcomes, and ensures that there \textcolor{black}{exists only one} version of treatment and that the potential outcomes for one region do not depend on treatment values of other regions (no interference). We believe this assumption is plausible in our study as we do not expect arrest rates in one region to depend on the police department practices of other regions. The second assumption is arguably stronger and states that for both the control and treated potential outcome point processes, the expected number of events that fall on one side of the boundary within a distance of $\delta$ is the same regardless of which side of the boundary is being looked at. This is a point process extension of local randomization assumptions used previously, and is needed because we need to use what happened under the precinct 0 side of the border to infer what would happen on the precinct 1 side of the border had they both been exposed to policing by the police department in precinct 0. As shown in Supplementary Materials Section 1, under these assumptions we can identify the effect of interest as
\begin{align*}
\theta(R_{\delta}) &= 2 E[Y(R_{\delta,1}) - Y(R_{\delta,0})],
\end{align*}
which is a fully observable quantity. Also of interest is $\tau(\boldsymbol{b})$, which represents the local treatment effect at location $\boldsymbol{b} \in \mathcal{B}$. To identify this effect, we need a point process extension of the conditional continuity assumptions used in regression discontinuity designs, which is given in assumption 2b. 

\noindent\textbf{Assumption 2b:} \textit{Continuity of potential outcome intensity surfaces}. $$\lim_{\boldsymbol{s} \to \boldsymbol{b}} \lambda^t(\boldsymbol{s}) = \lambda^t(\boldsymbol{b}) \quad \text{for } t=0,1. $$
This assumption guarantees that if we see a discontinuity at $\boldsymbol{b}$ in the observed intensity surfaces, then it can be attributed to an effect of the treatment, which also has a discontinuity there. Supplementary Materials Section 1 shows that $\tau(\boldsymbol{b})$ is identified under assumptions 1 and 2b as
\begin{align*}
    \tau(\boldsymbol{b}) = \lambda^1(\boldsymbol{b}) - \lambda^0(\boldsymbol{b}) = \lim_{\boldsymbol{s} \to \boldsymbol{b}^1} \lambda(\boldsymbol{s}) - \lim_{\boldsymbol{s} \to \boldsymbol{b}^0} \lambda(\boldsymbol{s}).
\end{align*}
Note here that we use the notation $\lim_{\boldsymbol{s} \to \boldsymbol{b}^1}$ to denote a limit that approaches the boundary location $\boldsymbol{b}$ from the precinct 1 side of the boundary, with analogous notation for precinct 0. Assumption 2a or 2b, depending on which estimand is being studied, is arguably the most crucial assumption needed for identification and the most likely one to not hold in practice. It could fail for instance if people decided where to live based on the geographic boundary, and therefore the two sides of the border would not be comparable with respect important characteristics that drive arrest rates, such as socioeconomic or criminological variables. Due to this, in Section \ref{sec: resamp} we detail a procedure to test the null hypothesis of no treatment effect that is robust to certain violations of these assumptions. Additionally, in Section \ref{sec:Covariates} and Supplementary Materials Section 2, we detail how additional covariate adjustment can be incorporated to weaken these assumptions. 

\subsection{Estimation of treatment effects}
\label{sec:estimation}

Before detailing our strategy for hypothesis testing that is robust to certain violations of assumption 2a or 2b, we must first discuss estimation strategies for the spatial regression discontinuity design with point process data. Estimation of $\theta(R_{\delta})$ is straightforward for a given choice of $\delta$, as a natural estimator is simply a plug-in estimator given by $\widehat{\theta}(R_{\delta}) = 2(Y(R_{\delta,1}) - Y(R_{\delta,0}))$, where $Y(R_{\delta,t})$ is the number of events on the precinct $t$ side of the boundary. In order to estimate $\tau(\boldsymbol{b}) = \lim_{\boldsymbol{s} \to \boldsymbol{b}^1} \lambda(\boldsymbol{s}) - \lim_{\boldsymbol{s} \to \boldsymbol{b}^0} \lambda(\boldsymbol{s})$, we fit two separate models, one for each of the two limits of interest. For $\lim_{\boldsymbol{s} \to \boldsymbol{b}^1} \lambda(\boldsymbol{s})$, we utilize only the data on the precinct 1 side of the boundary, and fit a model to estimate the intensity surface of the point process on that side of the boundary. We can then extrapolate this intensity surface to estimate the intensity surface at $\boldsymbol{b}$. Intensity surface estimation for this analysis utilizes the methodology implemented in the \texttt{R} package \texttt{spatstat} \citep{baddeley2005spatstat, baddeley2015spatial}. Specifically, let $W$ be a two-dimensional spatial window with which we define an intensity surface over. Further, let $\boldsymbol{l}_j$ represent a distinct spatial location of an observation within $W$ for $j \in \{1,\hdots, J\}$, and $m_j$ represents the weight associated with $\boldsymbol{l}_j$ (e.g., if there exists four observations located at $\boldsymbol{l}_j$, then $m_j = 4$). 
Then, at any given point $\boldsymbol{s}\in W$, the intensity surface value, $\lambda(\boldsymbol{s})$, is estimated using a fixed-bandwidth kernel estimate given by
$$\widehat{\lambda}(\boldsymbol{s}) = \frac{\sum_{j=1}^J \kappa(\boldsymbol{l}_j -\boldsymbol{s}) \cdot m_j}{e(\boldsymbol{s})}$$
where $e(\boldsymbol{s}) = \int_W \kappa(\boldsymbol{v}-\boldsymbol{s})\,d\boldsymbol{v}$ and $\kappa(\cdot)$ is the \textcolor{black}{kernel of an} isotropic Gaussian density function. The variance (i.e., bandwidth) for this density, denoted by $\sigma^2$, is what defines the spatial smoothness of the intensity surface. \textcolor{black}{The choice of $\sigma^2$ is one that empirically minimizes the mean squared error (MSE) of the estimator for the local intensity surface about $\vb*{s}$. The MSE for the estimator is derived from \cite{diggle1985kernel, berman1989estimating}, where it is assumed that the point-process is a stationary, isotropic Cox process, allowing the MSE to be written as a function of only the smoothness parameter $\sigma^2$.} Given their importance for all of the estimands considered, we discuss smoothness parameters such as $\sigma^2$ in more detail in Section \ref{sec:SmoothingParameters}.

\subsection{Resampling to obtain robust test of null treatment effect}
\label{sec: resamp}

Of major concern when using the regression discontinuity design is that the aforementioned assumptions do not hold. While assumption 1 is reasonable in many applications, assumptions 2a and 2b are relatively strong and can fail in certain scenarios. For simplicity, in this section we focus attention on $\theta(R_{\delta})$ and assumption 2a, but identical ideas hold for $\tau(\boldsymbol{b})$ and assumption 2b. Violations of assumption 2a are problematic as they can lead to bias in the estimated treatment effects and inflation of type I error rates, which could potentially be leading to the results seen in Figure \ref{fig:NaivePvalue}. For NYC in particular, this assumption would be violated if the communities on either side of the boundary are systematically different with respect to unmeasured variables that affect the potential outcome distributions. While we can reduce this by forcing $\delta$ to be as small as possible, neighborhoods in NYC can change drastically over short geographic distances. In other applications of GeoRDDs, units may choose to live on one side of the boundary due to the boundary itself, which can also violate this assumption. Our interest will be in testing the null hypothesis $H_0: \theta(R_{\delta}) = 0$, and our goal will be to create a hypothesis test that has valid type I error, even in the presence of certain violations of assumption 2a.

We develop a two step procedure to test this null hypothesis. The first step is to define a test statistic for this hypothesis, while the second step involves resampling new boundaries in NYC to estimate the distribution of this statistic under the null hypothesis of no precinct effect. 
For step one, we define 
\begin{align}
    Z = \left| Y(R_{\delta, 1}) - Y(R_{\delta, 0}) \right|, \label{eqn:TestStatistic}
\end{align}
which is the difference in the number of events between regions $R_{\delta, 1}$ and $R_{\delta, 0}$. Clearly, this test statistic will be large when there are differences in policing practices by police departments in precincts 1 and 0. One difficulty we must overcome is that this test statistic does not have a known distribution that can be used for inference.  A larger problem, however, is what happens when assumption 2a does not hold. Even if the distribution of this test statistic is known under the null hypothesis, violations of assumption 2a will lead to larger values of $Z$, and we must account for this to obtain valid inference.

Our goal is to estimate the null distribution of our test statistic, and we refer to the cumulative distribution function (CDF) of this distribution by $F_0$. To estimate this null distribution we can sample new precinct boundaries that behave similarly to the original precinct boundary of interest. The key difference is that these new boundaries, which we call \textit{null streets}, are fully contained within a single precinct and therefore have no precinct effect, i.e., $\theta(R_{\delta}) = 0$ by design. Fortunately, we have a very rich data set that includes information on all of NYC, not just at the boundaries of the precincts, and we can leverage this data set to find a large number of null streets. An illustration of this for one precinct can be found in Figure \ref{fig:NullStreets}, and a map showing streets across all of NYC can be found in Supplementary Materials Section 7.

\begin{figure}[!htb]
\spacingset{1}
    \centering
    \includegraphics[scale=0.2, trim={19cm 3cm 18cm 3cm},clip]{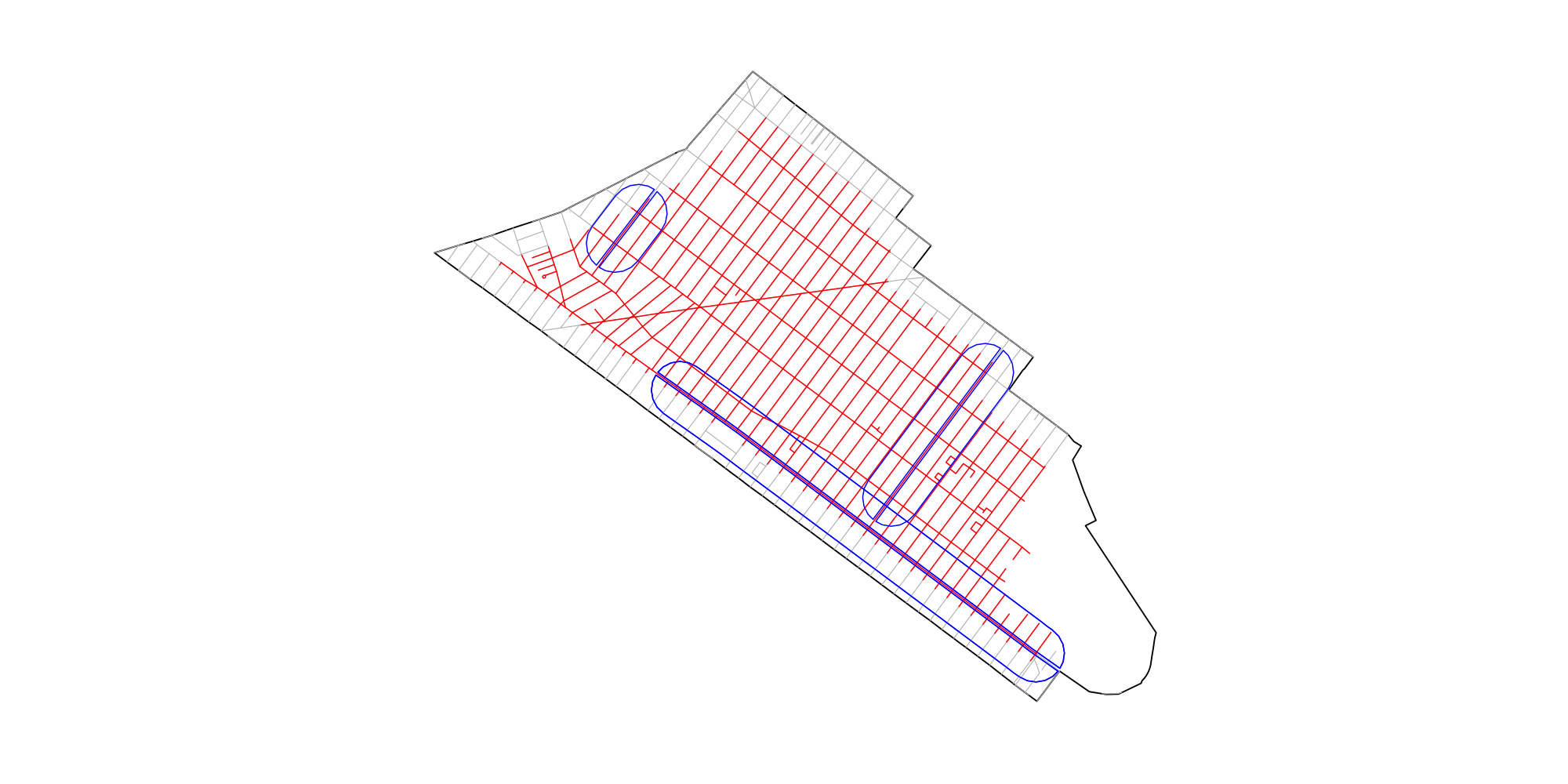}
        \caption{The red streets in Precinct 83 are ones that could potentially be used as null streets. The buffers (blue) are drawn around three potential null streets to illustrate how they meet the qualification for being completely contained in one precinct.}
    \label{fig:NullStreets}
\end{figure}

Assuming we can find a large number, $B$, of streets that are not near precinct boundaries, we can estimate the test statistic at each null street and use the distribution of these statistics as an estimate of $F_0$. Note that our procedure will be valid for any test statistic, though we will proceed with $Z$ from  (\ref{eqn:TestStatistic}).  We denote these test statistics by $Z^b$ for $b=1, \dots, B$.  We can then estimate the null distribution via $\widehat{F}_0(a) = \frac{1}{B} \sum_{b=1}^B \mathbf{1}(Z^b < a)$, which allows us to perform hypothesis testing. The intuition behind using this test to provide more robust hypothesis testing is that if assumption 2a does not hold at the boundary of interest, then it likely does not hold in other areas of NYC as well. For instance, there could be substantial spatial variability in individuals across NYC that changes far more locally than distances of $\delta$. We would not be able to account for this with observed covariates \textcolor{black}{that are available only at the census-tract level}, which is not sufficiently spatially resolved. However, it is likely that this variability is not unique to precinct boundaries, and that this variation also affects estimates at our resampled locations as well. By using these resampled locations, we are no longer relying on assumption 2a holding to obtain a valid hypothesis test, but rather a modified assumption at the resampled locations. To provide intuition for this, let us first simplify the test statistic to be $Z = Y(R_{\delta, 1}) - Y(R_{\delta, 0})$.
Under assumption 1, the mean of this difference can be written as
\begin{align*}
    E(Z) &= E[Y(R_{\delta, 1}) - Y(R_{\delta, 0})]\\
    &= \theta(R_{\delta}) + E[Y^0(R_{\delta, 1}) - Y^1(R_{\delta, 0})].
\end{align*}
If we further adopt assumption 2a, we have that $E(Z) = \theta(R_{\delta})/2$ and it is therefore zero under the null hypothesis of no precinct effect. If, however, assumption 2a is violated in the sense that $E\qty[Y^1(R_{\delta,1}) - Y^1(R_{\delta,0})] = E\qty[Y^0(R_{\delta,1}) - Y^0(R_{\delta,0})] = \mu$, we have that 
$$E(Z) = \theta(R_{\delta})/2 + \mu.$$
This shows that if there are structural differences on the two sides of the boundary with respect to the potential outcomes, this will inflate our test statistic. In particular, under the null hypothesis of no treatment effect, this expected difference becomes $\mu$ instead of zero, which can lead to inflated type I error \textcolor{black}{if this is not taken into account}. To describe when our procedure can lead to valid inference, we first define 
\begin{align*}
    U^t(\delta, \mathcal{B}) &= Y^t(R_{\delta,1}) - Y^t(R_{\delta,0}) \quad \text{for } t=0,1,
\end{align*}
where again $\mathcal{B}$ is used to denote the fact that these are at the boundary of interest. We also let $U^t(\delta, \mathcal{B}) \sim G$ for some distribution $G$ under the null hypothesis of no treatment effect. If we let $\mathcal{B}^*$ represent a null street boundary, then we make the following assumption:

\noindent \textbf{Assumption 3:} The null streets' test statistics match those of the precinct boundary locations under the null hypothesis, in the sense that $U^t(\delta, \mathcal{B}^*) \sim G$ for $t=0,1$. 

If this assumption holds and the null streets have similar levels of violations of assumption 2a, then the test statistics found at the null streets should approximate the true null distribution $F_0$ and we can construct rejection regions for our test using the relevant quantiles of the estimated null distribution. Assumption 2a makes the restrictive assumption that $E(U^t(\delta, \mathcal{B})) = 0$, while assumption 3 allows for violations in the sense that $E(U^t(\delta, \mathcal{B})) = \mu$ as long as we can find null streets with similar violations. We see in Section \ref{sec:theory} that this is actually a stronger assumption than what is required for valid inference. If the null streets have larger violations of assumption 2a than at the precinct boundaries, then we should obtain valid, albeit conservative, inference. Note that while we argued for this procedure for testing $\theta(R_{\delta}) = 0$, the same ideas hold for testing whether $\tau(\boldsymbol{b}) = 0$. We would simply need to change our test statistic  to be $Z = \qty|\lim_{\boldsymbol{s} \to \boldsymbol{b}^1} \widehat{\lambda}(\boldsymbol{b}) - \lim_{\boldsymbol{s} \to \boldsymbol{b}^0} \widehat{\lambda}(\boldsymbol{b})|$, and all other ideas remain unchanged.

\subsection{Choosing smoothing parameters}
\label{sec:SmoothingParameters}

\textcolor{black}{Note that for both estimands $\theta(R_{\delta})$ and $\tau(\boldsymbol{b})$, there exist a parameter that governs how localized estimation is over space. For $\theta(R_{\delta})$, the distance $\delta$ dictates a bias-variance trade-off for our estimation strategy because decreasing $\delta$ makes assumption 2a more plausible and reduces bias in estimation of the causal effect, however, it reduces the amount of data we have to estimate the treatment effect, thereby increasing variability. For $\tau(\boldsymbol{b})$, the spatial smoothing parameter $\sigma^2$ dictates a similar bias-variance tradeoff where a smoother intensity surface, $\lambda(\boldsymbol{b})$, will have lower variability, but may induce bias by using information too far from the boundary of interest.}

Related issues arise in standard regression discontinuity designs or GeoRDDs that utilize local linear regression for estimation, which contains a crucial bandwidth parameter. Selection of the bandwidth parameter has seen significant attention in the regression discontinuity design literature with a focus on finding optimal (in terms of MSE) bandwidth parameters \citep{imbens2012optimal} or performing inference in a way that accounts for bias in treatment effect estimation \citep{calonico2014robust, calonico2014robustb}. One key issue in this literature is that constructing valid confidence intervals when using cross-validation or MSE-optimal choices of bandwidth parameters is difficult because of the asymptotic bias in these estimators caused by oversmoothing \citep{keele2015geographic}. One can attempt to undersmooth by choosing a smaller bandwidth than what is chosen by cross-validation, but the degree of undersmoothing is generally not known. Bias-corrected confidence intervals have been developed in standard regression discontinuity designs, but this theory has not been developed for the point process setting seen here. Fortunately, our resampling procedure described in Section \ref{sec: resamp} helps resolve some of these issues. 
\textcolor{black}{For estimation of $\tau(\boldsymbol{b})$, we recommend choosing the smoothing parameter $\sigma^2$ that minimizes the MSE of the intensity surface estimator.}
This should lead to good estimates of the intensity surfaces that balance competing interests of using enough data, while also focusing in a close window around the boundary. 
\textcolor{black}{In our case, the MSE-optimal $\sigma^2$ is determined by a numerical minimization of the MSE estimate defined in \cite{diggle1985kernel} and \cite{berman1989estimating}. Although traditional cross-validation through resampling (e.g., leave-one-out cross-validation) could be used as another model-selection strategy, we find that for the size of our data it is computationally prohibitive. Additionally, both the spatial dependencies in the data as well as the fact that our data consist of only \textit{one} realization of the point-process, makes the independence assumption for cross-validation unreasonable. Hence, an area to explore in the future is applying novel cross-validation strategies that account for spatial dependencies, such as the work of \cite{cronie2018non} and \cite{cronie2024cross}, to improve the choice of $\sigma^2$ thus improving the overall intensity surface estimate. In any case, while the possibility of undersmoothing or oversmoothing exists, it is expected that these issues occur at both the boundaries of interest and the null streets in a similar manner, thus leading to valid hypothesis tests.}

Finding an optimal choice for $\delta$ to be used when studying $\theta(R_{\delta})$ is less clear, however, because there is no natural way to perform cross-validation with respect to $\delta$. Nonetheless, we can provide general guidance for choice of this parameter. 
A general rule of thumb is to set $\delta$ to be a small value, which has two benefits in our framework. For one, it makes assumption 2a more plausible than larger values of $\delta$. It also has an advantage with respect to the resampling procedure described in the previous section, which is that smaller values of $\delta$ will have more available null streets to select from, which can lead to null streets being more similar to the precinct boundaries of interest. This is because null streets, and their corresponding buffer regions, must be fully contained within a single precinct, but this becomes less likely as the size of the buffer region grows. \textcolor{black}{One approach we recommend to determine what a ``small value'' for $\delta$ means in any application is to first consider the spatial surface over which the point-process lies. In our case, we know all observations fall along street lines. Hence, we can look at the distances between neighboring streets (as quantified by the distance between street midpoints) and compute the minimum distance between each street and its neighbors within each precinct. We can then set $\delta$ to be the 95\%-quantile of the distribution of the minimum distance from streets to their nearest neighboring street. In our application in NYC, the 95\%-quantile is 273.25 feet, and therefore we recommend $\delta = 300$ as it is greater than most minimum distances between neighboring streets, while still small enough to make assumption 2a more plausible. Another} way in which data can be used to select $\delta$ is if additional information on a negative control variable is available, such as in our study of tree locations in NYC. One can apply the resampling procedure for all possible values of $\delta$ and choose the value of $\delta$ that leads to the desired type I error rate.

In general, however, we recommend finding a small value of $\delta$ as described above and then performing inference for a range of small to moderate values of $\delta$ \textcolor{black}{(we consider $\delta \in \{300, 400, \hdots, 1000\}$)}. This assesses whether results are consistent across these values, which would increase belief in the overall findings. Additionally, our testing procedure should be fairly robust to the choice of $\delta$ (or $\sigma^2$). In the NYC policing analysis in Section \ref{sec:NYCanalysis}, we apply our procedure for a range of smoothness parameters and find relatively consistent results across all values explored.

\subsection{The role of covariates}
\label{sec:Covariates}
As with nearly any observational study aiming to study causal effects, we must discuss the different ways in which covariates are accounted for. This is particularly important in the present setting, as there are multiple manners in which covariates can be included in our analysis, and it is important to distinguish among these. There are two distinct places that covariates can be incorporated: (1) the identification assumptions and corresponding estimation strategy described in Sections \ref{sec:Identification} and \ref{sec:estimation}, and (2) utilizing covariates to find the best null streets for estimating the null distribution in Section \ref{sec: resamp}. While commonalities exist across these two aspects of our proposed procedure, key distinctions remain which are worth spelling out. 

\subsubsection{\textcolor{black}{Effect on identification assumptions}}
Before discussing how covariates can be explicitly incorporated into the proposed procedure, we must also emphasize that regression discontinuity designs are useful, and so widely used, because they implicitly adjust for important confounding variables by design. In the context of spatial regression discontinuity designs, if important confounding variables are expected to vary smoothly across space, then the GeoRDD eliminates issues stemming from these variables by estimating treatment effects at the boundary. 
If the confounding factors are continuous at the boundary of interest, then the potential intensity surface will be continuous as well, and the GeoRDD can identify causal effects even without explicit adjustment for these variables. This logic has led to the regression discontinuity design being used in a variety of settings without the additional adjustment of covariates. This can be violated, however, in certain settings, such as at county or state lines, where important variables might change drastically at the boundary, as different counties have better schools, childcare options, or other factors influencing who ultimately decides to live there. In the current context of police precinct boundaries, this is expected to be less of an issue as these boundaries do not typically coincide with other important government boundaries that influence the type of people living in each area. \textcolor{black}{Notably, a recent sociological study titled \textit{Upsold} investigates ``consumers’ preferences and decision-making in the context of purchasing homes'' \citep{besbris2020upsold}, and among the NYC home-buyers, police precinct was \textit{not} a factor considered. Similar conclusions were found in a separate, recent book titled \textit{Race Brokers} \citep{korver2021race}. It is still possible, however, that police precincts could align with other sub-municipal boundaries that do affect where people live. We study this in Supplementary Materials Section 9 where we find that police precincts do not generally align with other important boundaries in NYC, which helps justify the assumption that precincts do not typically influence people's residential choices in NYC. Despite our justification above, we acknowledge that this is still an unverifiable assumption, which could affect the validity of our results if violated.}

If there are still concerns about differences in the distribution of important confounding factors on the two sides of the boundary of interest, then additional covariate adjustment can be incorporated to remove these differences, which increases the plausibility of the GeoRDD. In this setting, when referring to covariates, we are referring to spatial covariates that describe features of the geographic areas examined. For instance, one might be interested in adjusting for socioeconomic status if it is thought that socioeconomic status differs drastically on one side of the boundary compared with the other. The identifying assumptions for the point process GeoRDD can be relaxed to hold conditionally on observed covariates. We describe these extensions, the identification of causal effects incorporating covariates, and corresponding estimators in detail in Supplementary Materials Section 2. We do not implement this explicit covariate adjustment in our NYC policing analysis in Section \ref{sec:NYCanalysis}, because the only covariates available to us are United States Census variables, which are constant within Census blocks, and therefore are not spatially varying enough to assist in our analysis. 

\subsubsection{\textcolor{black}{Finding null streets}} \label{subsubsec:null}
The second way in which covariates influence our testing strategy, which we do implement in Sections \textcolor{black}{\ref{sec:sim} and \ref{sec:NYCanalysis}}, is in the selection of null streets. Assumption 3 states that the distribution of test statistics at the null streets and the precinct boundaries of interest should be equal under the null hypothesis of no police precinct effect. The distribution of the test statistics can depend on a number of covariates, however, and we need to incorporate these into the selection of which streets to use when estimating the null distribution for our test statistics. These covariates may be potential confounding factors, such as socioeconomic status, but they need not be. For instance, the size of the null street is a potentially important factor to use when selecting null streets as small streets will have far fewer data points than large streets, and will subsequently have more variability in their corresponding test statistics. \textcolor{black}{Given the importance of null streets, we now detail the steps taken to find adequate null streets to be used in the estimation of the null distribution for a given precinct boundary's test statistic.}

\textcolor{black}{The first step in determining which null streets are ``similar'' to the boundary of interest is to choose features in the data that potentially have an affect on the distribution of the test statistic being computed. This step is largely application dependent and Table \ref{tab:nullCov} shows a list of the different covariates used to find null streets for each of the three analyses we perform. Note that for a given boundary, there exist two values for each covariate; one on each side of the boundary. As a general guideline, the choice of covariate is based on both expert knowledge as well as specific features of the data application. For the negative control analysis in Section \ref{sec:NegControl}, we know the data are observed only on streets, so including relevant street information as a covariate is justifiable. For the simulation in Section \ref{sec:sim}, we randomly place data across the entire geographic domain, and the data are not restricted to street locations. Therefore the total area/size of the buffer region is now a more relevant covariate to include. Finally, in the NYC policing analysis in Section \ref{sec:NYCanalysis}, we use the amount of crime as a covariate because it is expected that areas with higher amounts of crime will have more arrests, irrespective of which police precinct governs that area. While each covariate we included in order to find null streets has a practical justification based on subject matter expertise, Supplementary Materials Section 4 presents empirical justifications for their use by showing the association of these covariates with the magnitude and variability of test statistics under the null hypothesis. Additionally, while we use a single numeric feature to find null streets in each analysis, this can be extended to using multiple covariates, if it is expected that these additional features impact the distribution of the test statistic.}
\begin{table}
    \centering
    \begin{tabular}{|l||l| }
    \hline
    Data application & Covariate/Feature\\
    \hline
    Simulation & buffer region area\\
    Negative Control  & street length in buffer region\\
    Arrest Data & crime locations in buffer region\\
    \hline
    \end{tabular}
    \caption{The choice of covariate used to find null streets for each application. Note that these covariates are numeric summaries and are distinct for \textit{each side} of a given boundary.}
    \label{tab:nullCov}
\end{table}

\textcolor{black}{Now, for a given choice of covariate, we can formally describe how it is used to find adequate null streets. Let $\mathcal{M}_i$ denote the set of null streets for precinct boundary $\mathcal{B}_i, i \in \{1,2,\dots, 144\}$. Further, let $t^{\text{sum}}(\cdot)$ denote the sum of the covariate values from both sides of the boundary (e.g., the total amount of crime), and let $t^{\text{ratio}}(\cdot) \geq 1$ denote the ratio of the covariate values from each side of the boundary (e.g., the ratio of the amount of crime on the two sides of the boundary). Then, a street $\vb*{m}$ is considered a null street for $\mathcal{B}_i$ (i.e., $\vb*{m}\in \mathcal{M}_i$) if both $c\cdot t^{\text{sum}}(\mathcal{B}_i) < t^{\text{sum}}(\vb*{m}) < (1/c)\cdot t^{\text{sum}}(\mathcal{B}_i)$ and $c\cdot t^{\text{ratio}}(\mathcal{B}_i) < t^{\text{ratio}}(\vb*{m}) < (1/c)\cdot t^{\text{ratio}}(\mathcal{B}_i)$. Here $c\in(0,1)$ measures the degree of similarity that a street must have to the boundary of interest to be chosen as a null street. Values of $c$ near 1 ensure null streets are very similar to boundary $\mathcal{B}_i$ with respect to the variable being considered, though potentially at a cost of reducing the number of available null streets. Alternatively, a value of $c$ close to 0 means null streets are less similar to the boundary $\mathcal{B}_i$, but there will exist a large number of available null streets. Section \ref{sec:MatchesNYC} offers further commentary on choosing the value of $c$ to use in our motivating application. Once a value of $c$ is chosen, the null distribution for the test statistic at $\mathcal{B}_i$ is constructed by computing the test statistic for every $\vb*{m}\in \mathcal{M}_i$. Lastly, the reason for using $t^{\text{sum}}(\cdot)$ and $t^{\text{ratio}}(\cdot)$ can best be explained in the context of the real data analysis of Section \ref{sec:NYCanalysis}. We choose null streets based on the total amount of crime because areas with more people and more events are likely to have lower variability in their corresponding test statistics. We also use the ratio of crimes on the two sides of the boundary, because crime likely correlates with many important confounding factors that we do not measure, and finding null streets based on this ratio helps ensure that the null streets have similar violations of assumptions 2a or 2b, which makes assumption 3 more plausible. Given the importance of this choice of covariate to our resampling strategy, we study it in further detail in the following section.}

\subsection{Theoretical insights for resampling procedure}
\label{sec:theory}
In this section, we examine when the resampling procedure will provide valid inference, and provide guidance for choosing the number of null streets $B$, which is a critical choice for both statistical validity and power of the proposed hypothesis tests. Larger values of $B$ should lead to more efficient estimates of the unknown CDF $F_0$. However, increasing $B$ also may lead to using null streets that do not closely match the boundary of interest and assumption 3 will not be satisfied. For the rest of this section, assume a fixed buffer width $\delta$, though all results will hold regardless of the chosen buffer width if assumption 3 is satisfied. Let $Z_i$ represent the test statistic at precinct boundary $i$. Further, let $Z_i^{(b)}$ represent a resampled test statistic for precinct boundary $i$ where $b \in \{1, 2, ..., B\}$. Next, define $X_i$ to be characteristics associated with precinct boundary $i$ and similarly, $X_i^{(b)}$ denotes characteristics for null street $b$. For simplicity, we let $X_i$ and $X_i^{(b)}$ be univariate here, but the same ideas hold for a vector of covariates. Assume under the null hypothesis that $P(Z\leq z \mid X=x) = F(x,z)$ is the cumulative distribution function of the test statistic of interest, and $f(x,z)$ represents the corresponding density function. Note that we are assuming that the distribution of the test statistic depends on characteristics $X$. Potential characteristics in NYC are the length of the border $\mathcal{B}$ or the ratio of crime rates on the two sides of the border. \textcolor{black}{As noted in Section \ref{subsubsec:null},} the distribution of the test statistic in NYC indeed depends heavily on characteristics such as these. For this reason, we want to find null streets with similar values of these characteristics, i.e., $X_i^{(b)} \approx X_i$ for $b=1, \dots, B$ so that assumption 3 holds, though we quantify this idea more rigorously in what follows. 

Our first goal of the resampling procedure is to obtain statistical validity and maintain type I error control at rate $\alpha$ for our hypothesis test. We can define the true $1-\alpha$ quantile of $Z_i$ under the null hypothesis as $Q_{1-\alpha}$. Our corresponding estimate of this quantile is given by 
$$\widehat{Q}_{1-\alpha} = \min \bigg\{ q: \widehat{F}(X_i, q) \geq 1 - \alpha \bigg\},$$
where our estimate of the CDF is given by
$$\widehat{F}(X_i, q) = \frac{1}{B}\sum_{b=1}^B \mathbf{1}(Z_i^{(b)} \leq q).$$
We use this estimate of the CDF throughout this section, though kernel smoothing can be used to improve estimation of the CDF when $B$ is small. In Supplementary Materials Section 3, we show that the type I error of our procedure is given by 
$$P(\text{reject } H_0 \; | \; H_0) = 1 - E_{\widehat{Q}} \qty[F(X_i, \widehat{Q}_{1-\alpha})].$$
This result implies that we can obtain type I error control at level $\alpha$ if 
$$E_{\widehat Q} \qty[F(X_i, \widehat{Q}_{1-\alpha})] \geq 1- \alpha = F(X, Q_{1-\alpha}).$$
This shows that validity does not rely on an unbiased, or even conservative, estimate of $Q_{1 - \alpha}$. Rather we need our estimated quantiles $\widehat{Q}_{1-\alpha}$ to be such that on average, the true CDF at our estimated quantiles is above $1-\alpha$. Even if we have an unbiased estimator of $Q_{1 - \alpha}$, if it has excessive variance, then it might not satisfy the condition above and will lead to anti-conservative inference. To gain further intuition into this, we study the properties of our estimator of the CDF, denoted by $\widehat{F}(X_i, q)$. First, we can look at the mean of this estimator, which we show in Supplementary Materials Section 3 can be approximated as follows:
\begin{align*}
    E[\widehat{F}(X_i, q)] \approx F(X_i, q) &+ \frac{d}{dX_i} F(X_i, q) \cdot E(X_i^{(b)} - X_i) \\
    & + \frac{d^2}{dX_i^2} F(X_i, q) \cdot E\qty[(X_i^{(b)} - X_i)^2].
\end{align*}
This shows that the bias of the estimator is a function of how close the null street covariates $X_i^{(b)}$ are to $X_i$. Therefore, finding null streets that have similar characteristics as the precinct boundary of interest is of crucial importance. Of course, if $X_i$ does not affect the distribution of the test statistic, then $\frac{d}{dX_i} F(X_i, q) = \frac{d^2}{dX_i^2} F(X_i, q) = 0$ and we have no bias regardless of how similar the null streets are. Also of importance is the variance of this estimator, which we show in Supplementary Materials Section 3 is approximated by
\begin{align*}
    \text{Var}[\widehat{F}(X_i, q)] &\approx \frac{1}{B} \Bigg\{E\qty[F(X_i^{(b)}, q) \cdot (1-F(X_i^{(b)}, q))]\\
    &\quad \quad+ \qty(\frac{d}{dX_i}F(X_i, q))^2 \cdot Var(X_i^{(b)})\Bigg\}.
\end{align*}
One would expect that the variance generally decreases as we increase the number of null streets, $B$. However, this shows an important feature of the resampling procedure, which is that the variance need not necessarily go down as we increase $B$. This is because increasing $B$ can also increase $\text{Var}(X_i^{(b)})$ by including null streets with very different values of $X_i^{(b)}$, which leads to an increase in the overall variance. These two results show that there is a trade-off involved when choosing $B$. On one hand we want to increase $B$ to decrease variability in $\widehat{F}(X_i, q)$. On the other hand, we want to keep $B$ small enough so that the null streets are closely aligned with the precinct boundary of interest in the sense that $X_i^{(b)} \approx X_i$, which reduces bias, and potentially reduces the variance of the estimate of the CDF. We discuss this trade-off in the context of the NYC data in Section \ref{sec:MatchesNYC}.

\subsection{Global test of variation by precinct}
\label{sec:Global}

So far we have focused on performing a hypothesis test at a single border (namely between precincts 0 and 1), but there exists many such precinct borders in NYC.
While there is interest in knowing whether any two bordering precincts have differential arresting practices, also of interest is whether there is any variation across all NYC police precincts. In this setting we might wish to test whether the arrest rates differ by any precinct in NYC. It is difficult to compare any two precincts that are not bordering each other as we would not be able to focus on the border between these two precincts and therefore cannot utilize the GeoRDD. For this reason we restrict our attention to assessing whether any bordering precincts have differential precinct practices. Focusing first on the local average treatment effect within $\delta$ of the boundary, let $\theta(R_{\delta}^{(i)}) = E(Y^1(R_{\delta}^{(i)}) - Y^0(R_{\delta}^{(i)}))$ be the local treatment effect at boundary $i$, thus leading to the following hypothesis:
\begin{align*}
    & H_0: \theta(R_{\delta}^{(i)}) = 0 \text{ for } i=1, \dots, M \\
    & H_a: \theta(R_{\delta}^{(i)}) \neq 0 \text{ for at least one } i.
\end{align*}
Note that $R_\delta^{(i)}$ is defined precisely as $R_\delta$ is defined in Section \ref{subsec:sbe} except $i = 1, 2, \hdots, M$ specifies the exact boundary of interest.
To perform this hypothesis test, we use a test statistic given by 
$$\bar{Z} = \frac{1}{M} \sum_{i = 1}^M \ Z_i = \frac{1}{M} \sum_{i = 1}^M \left| Y(R_{\delta, 1}^{(i)}) - Y(R_{\delta, 0}^{(i)}) \right|.$$
Larger values of this test statistic provide additional evidence against the null hypothesis. We can use the same resampling procedure described in Sections \ref{sec: resamp} and \ref{sec:theory} in order to perform inference using this test statistic. We approximate the distribution of $\bar{Z}$ under the null hypothesis using the empirical distribution of $\bar{Z}^b$ for $b=1, \dots, B$. Note that while we focused on the local average treatment effect within $\delta$ of the precinct boundaries, the same ideas would apply for $\tau$ defined in \eqref{eqn:WLATE}. Specifically, we could use the resampling procedure to test
\begin{align*}
    & H_0: \tau_i = 0 \text{ for } i=1, \dots, M \\
    & H_a: \tau_i \neq 0 \text{ for at least one } i.
\end{align*}
where $\tau_i$ is the estimand defined in (\ref{eqn:WLATE}) applied to precinct boundary $i$. Similarly, the test statistics $Z_i$ would be updated to be estimators of $\tau_i$, which are defined in Section \ref{sec:estimation}. Note that while the test statistic is an average across all precinct boundaries, another justifiable test statistic would be to use $\max_i Z_i$, which is analogous to using a minimum p-value over all hypothesis tests \citep{tippett1931methods}. This statistic may have more power if only a small subset of the precinct boundaries have an effect of police precincts. 

\section{Simulation study} \label{sec:sim}

Here we assess the performance of the proposed approach to testing in the GeoRDD using simulated outcome data across NYC. We generate data from four scenarios to evaluate performance in a wide range of plausible settings. In each scenario, we generate 1000 data sets. In each data set, we first generate the intensity surface $\lambda(\cdot)$ of the point process across the surface of NYC. Counts of outcomes within any particular region $R$, such as the area around a precinct boundary, are then drawn from a Poisson distribution with mean given by $\Lambda(R) = \int_R \lambda(\boldsymbol{s}) d \boldsymbol{s}$. Similar to the negative control analysis, simply doing a binomial test to compare the number of simulated counts on either side of a boundary leads to invalid results and inflated type I error rates. With this in mind, we run our proposed procedure on each of the simulated data sets and evaluate the probability that the null hypothesis of no precinct effect is rejected. For the individual precinct boundary tests, results are averaged over all 1000 data sets and all 144 precinct boundaries in NYC. For the global test, only one test is run for each simulated data set, and results are averaged over 1000 simulations. 

\subsection{Surface construction}
\label{subsec:surfcon}
\begin{figure}[!htb]
\spacingset{1}
    \centering
    \includegraphics[width = \linewidth]{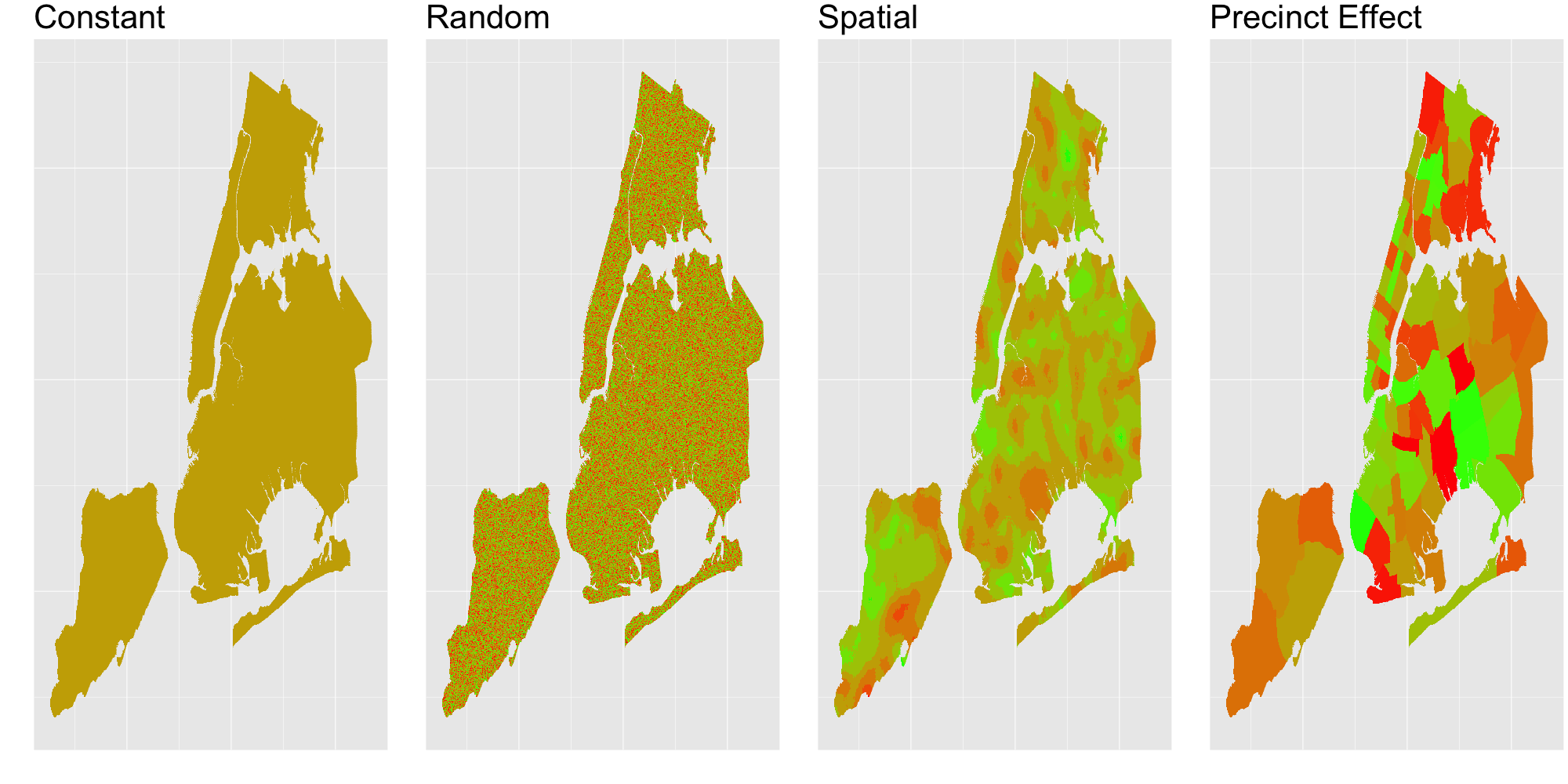}
    \vspace{-.1in}
    \caption{Heat maps representing one realization of an intensity surface for the outcomes.}
    \label{fig:heat_maps}
    \vspace{-0.1in}
\end{figure}

We refer to the four different underlying intensity surfaces of the observed outcomes for the NYC landscape as (1) \textit{Constant}, (2) \textit{Random}, (3) \textit{Spatial}, and (4) \textit{Precinct Effect}. Heat maps of one realization for each of the surfaces are shown in Figure \ref{fig:heat_maps} where the dark red areas represent higher values of the outcome. The Constant, Random, and Spatial surfaces represent situations with no precinct effect, and therefore the null hypothesis of no precinct effect is true. The Random and Spatial surfaces, however, represent situations where standard regression discontinuity designs might fail because there will likely be more counts on one side of the precinct boundary than the other due to randomness or spatial variation in the surfaces not driven by any precinct effect. These are intended to represent realistic situations in NYC such as crime hotspots or spatial correlation in crime levels that can lead to differential counts of outcomes in one precinct than another that is not attributable to the precincts themselves. In these situations, our goal is to maintain type I error control at level $\alpha$ despite these differential counts. The Precinct Effect surface, however, has clear precinct effects and we want to assess the power to detect these differences.

\subsection{Type I error control and power}
\label{subsec:results}
\begin{table}
    \centering
    \begin{tabular}{|l||l|l|l|l| }
    \hline
    \multicolumn{5}{|c|}{Individual}\\
    \hline
    $\delta$ &Constant &Random &Spatial &Precinct \\
    \hline
    300 & 0.049 & 0.054 & 0.051 & 0.920\\
    400 & 0.048 & 0.054 & 0.053 & 0.935\\
    500 & 0.048 & 0.051 & 0.054 & 0.942\\
    600 & 0.050 & 0.053 & 0.063 & 0.953\\
    700 & 0.051 & 0.058 & 0.074 & 0.963\\
    800 & 0.050 & 0.057 & 0.080 & 0.967\\
    900 & 0.051 & 0.057 & 0.082 & 0.970\\
    1000& 0.052 & 0.059 & 0.089 & 0.973\\
    \hline
    \hline
    \multicolumn{5}{|c|}{Global ($\max_i Z_i$; $\bar{Z}$)}\\
    \hline
    $\delta$ &Constant &Random &Spatial &Precinct \\
    \hline
    300 & 0.038; 0.040 & 0.032; 0.041 & 0.017; 0.024 & 1.000; 1.000\\
    400 & 0.048; 0.024 & 0.028; 0.034 & 0.019; 0.030 & 1.000; 1.000\\
    500 & 0.040; 0.028 & 0.031; 0.022 & 0.034; 0.045 & 1.000; 1.000\\
    600 & 0.048; 0.032 & 0.041; 0.034 & 0.043; 0.108 & 1.000; 1.000\\
    700 & 0.044; 0.035 & 0.040; 0.056 & 0.061; 0.231 & 1.000; 1.000\\
    800 & 0.045; 0.038 & 0.031; 0.045 & 0.077; 0.297 & 1.000; 1.000\\
    900 & 0.035; 0.045 & 0.048; 0.049 & 0.091; 0.348 & 1.000; 1.000\\
    1000& 0.056; 0.053 & 0.038; 0.054 & 0.111; 0.408 & 1.000; 1.000\\
    \hline
    \end{tabular}
    \vspace{0.1in}
    \caption{Probability of rejecting the null hypothesis across the four simulation scenarios and differing buffer widths. The top half of the table corresponds to hypothesis tests at individual precinct boundaries, while the bottom half corresponds to the global test of variation across NYC using two different test statistics.}
    \label{tab:simP}
    \vspace{-0.25in}
\end{table}
The results from all simulations can be found in Table \ref{tab:simP}, which shows the percentage of rejected hypothesis tests for both the individual tests and global tests, respectively. Note that these results are for hypotheses in terms of $\theta(R_\delta)$, not $\tau(\boldsymbol{b})$. Estimation of $\tau(\boldsymbol{b})$ requires \textcolor{black}{finding the MSE-optimal smoothing parameter} for the intensity surface estimation at all boundaries of interest, including null streets, and is therefore computationally prohibitive to run on such a large scale over 1000 simulations. We see that for the Constant, Random, and Spatial surfaces, we are able to recover $\alpha = 0.05$ type I error rates. The Spatial surface is somewhat more challenging and leads to slightly inflated type I error rates for larger buffer widths, while maintaining type I error control at smaller buffer widths. In terms of the global test, both statistics perform relatively well, though the maximum statistic is somewhat more robust with smaller type I error rates in the Spatial surface setting. Overall, this shows that the proposed approach is indeed able to provide valid inference even in settings with differential outcome levels on either side of the boundary, i.e., when assumption 2a is violated. \textcolor{black}{In this setting, violations of assumption 2a occur in a similar manner across the city, and therefore assumption 3 holds and we obtain valid inference.} In the Precinct Effect scenario, the null hypothesis is not true, and we see that our approach has high power to detect these differences across precincts. The power is slightly below 1 for the individual tests and this is because some precincts have very small differences in counts from their neighboring precincts. The global test does not suffer from this issue as it uses either the average or maximum of the test statistics over all precinct boundaries, leading to a power of 1. \textcolor{black}{Additionally, in Supplementary Materials Section 8 we run additional simulation studies with reduced sample sizes and obtain similar results.} 

\section{Analysis of precinct by precinct arrest rates}
\label{sec:NYCanalysis}

Here we analyze the NYC arrest data to estimate the degree of variation in policing across the city as well as whether there are significant differences between individual precincts with regards to their arresting practices. We first discuss our strategy for finding null streets in NYC and use our procedure to test the null hypothesis of no precinct effects in the negative control data. Then, we use our procedure to test for precinct-specific effects and global variation in policing with respect to arrest rates. Given that our outcome of interest is arrest rates and not raw totals of arrests, we scale all results by the number of crimes in the corresponding area. For $\theta(R_{\delta})$ this is done by dividing all arrest counts by the number of crimes in the same area, and for $\tau$ this is done by dividing intensity surfaces for arrests by the intensity surfaces for crimes.

For null streets and the negative control analysis, if an event occurs directly on the boundary between the two sides of interest, that point is randomly assigned to one side of the boundary. For the main analysis of police precincts, we have information on the precinct of the arresting officer, and we use this to assign a precinct to observations that fall directly on the border. Additionally, throughout this section, we present results for differing spatial smoothness values. The spatial smoothness level is determined by first using the \textcolor{black}{MSE-optimal estimate} of $\sigma^2$ as in Section \ref{sec:SmoothingParameters}, then scaling $\hat{\sigma}^2$ by some factor (``smoothing multiplier'') to vary the smoothness. Also, note that for constructing intensity surfaces, we use observations that fall within a specified radius (``region size'') from the boundary of interest. We focus on a region size of 600 feet here, but ran the same analyses for a range of region sizes and found very similar results. Lastly, all results presented in this section are for hypotheses with respect to $\tau$. To see analyses targeting $\theta(R_{\delta})$, as well as additional results from the analyses for $\tau$, see Supplementary Materials Sections 5 and 6.

\subsection{\textcolor{black}{Quantifying the similarity of null streets}}
\label{sec:MatchesNYC}
\begin{figure}[!htb]
\spacingset{1}
    \centering
        \includegraphics[page = 2, trim={0 3.5in 0 0.05in},clip, width = 0.49\linewidth]{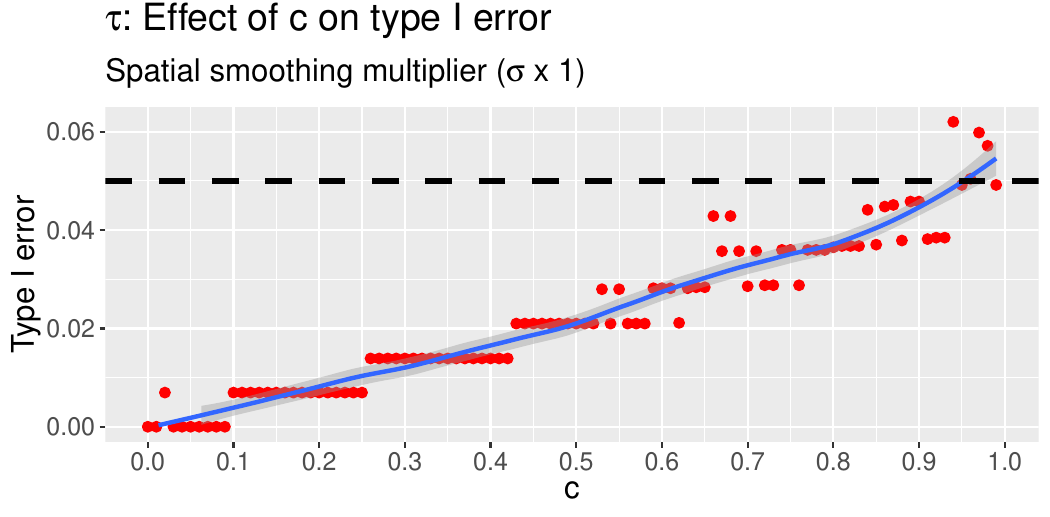}
        \vspace{-.1in}
        \caption{\textcolor{black}{The type I error as a function of $c$, a measure of how similar the null streets are to the precinct boundaries of interest.}}
    \label{fig:numMatch}
\end{figure}

\textcolor{black}{As discussed in Sections \ref{sec:Covariates} and \ref{sec:theory},} it is important to consider both which covariates to use when finding null streets and the number of null streets to use, as both have implications for type I error control. One feature of streets that can alter the distribution of the test statistics under the null hypothesis is the size of the street being considered. We expect larger streets with more crime to have less variability in their test statistics, while small streets with small counts on either side of the boundary to have much more variability. \textcolor{black}{In other words,} the null streets can be used to learn which covariates impact the test statistics and therefore should be incorporated in the final analysis. \textcolor{black}{Recall from Section \ref{subsubsec:null} that} $c\in(0,1)$ measures the degree of similarity that a street must have to the boundary of interest. \textcolor{black}{Figure \ref{fig:numMatch} illustrates the importance of $c$ as it displays the type I error for the negative control analysis of Section \ref{sec:NegControl} as a function of $c$.} We see that large values of $c$ lead to type I error rates close to the desired level $\alpha = 0.05$. However, as $c$ approaches 0 ($c=0$ means all streets are considered null streets and covariates are not incorporated) the type I error approaches zero. Smaller values of $c$ lead to less similarity between the null streets and the precinct boundary of interest, which in this case leads to overly conservative inference. For this reason, we proceed with $c = 0.9$ moving forward to ensure our procedure has well-calibrated type I error and there are sufficient numbers of null streets to estimate the null distribution.

\subsection{Negative control analysis}
\label{sec:NegControl}
\begin{figure}[!htb]
\spacingset{1}
    \centering
    \includegraphics[width=0.49\linewidth]{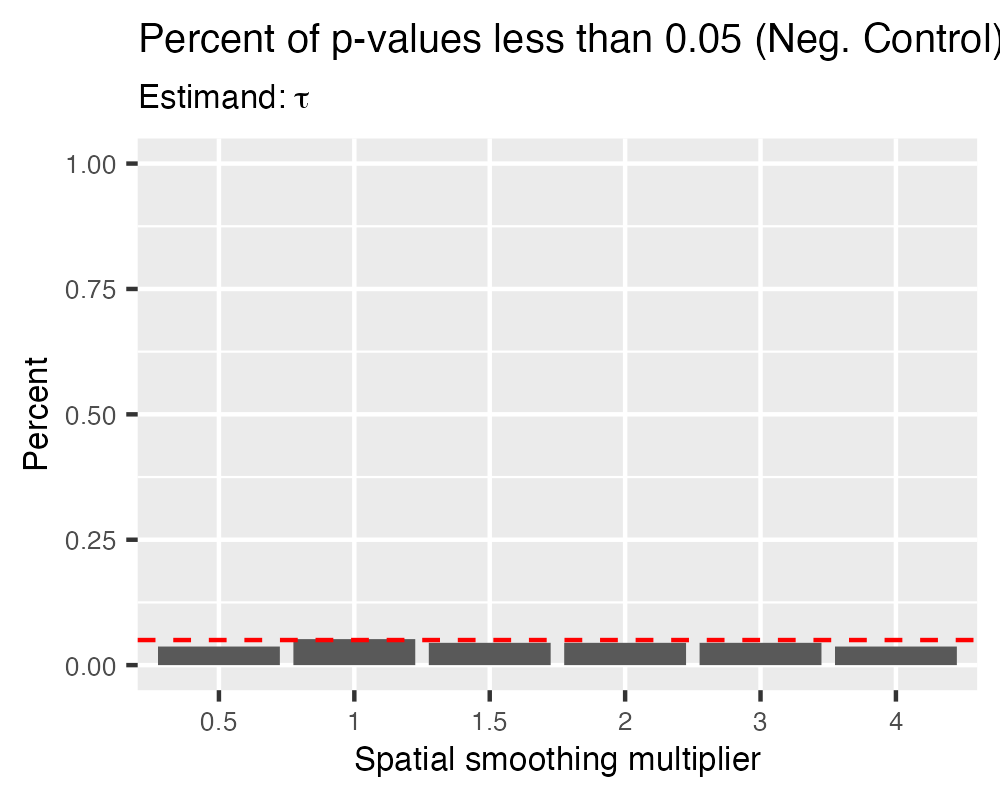}
    \includegraphics[width=0.49\linewidth]{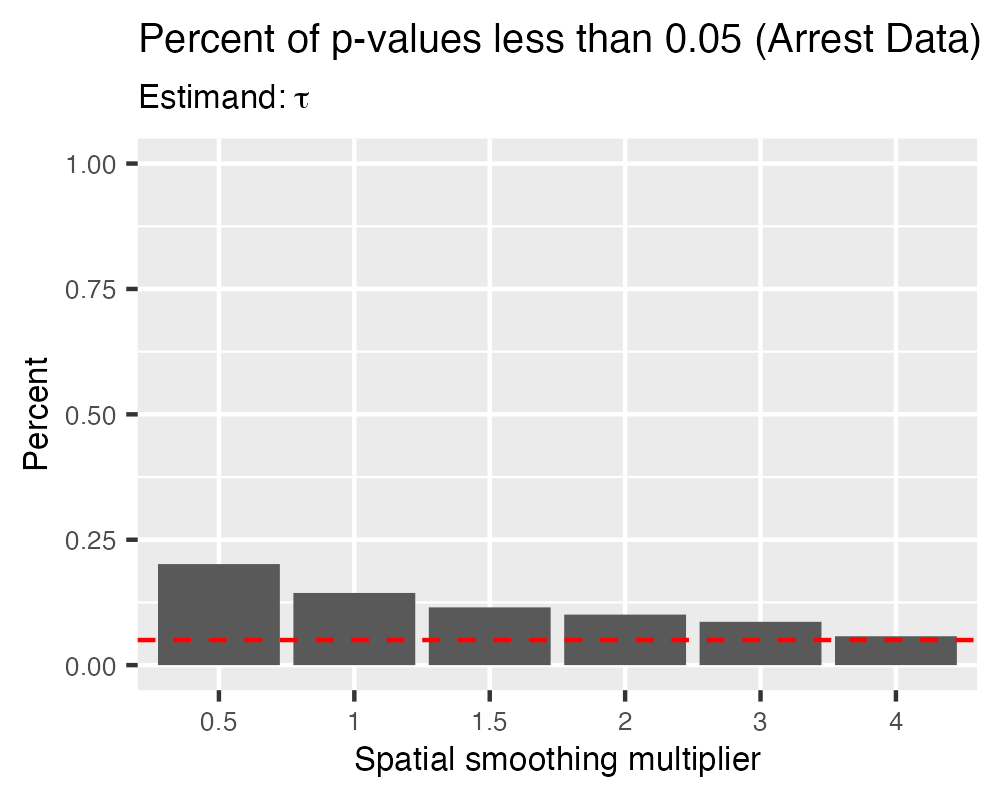}
    \vspace{-.1in}
    \caption{The percentage of p-values across the 144 boundaries that are less than 0.05 using test statistics calculated from the constructed intensity surfaces and the proposed resampling procedure. The top panel provides the results for the negative control analysis, and the bottom panel corresponds to the arrest data analysis. We explored various smoothing parameter values with larger values corresponding to smoother intensity surfaces.}
    \label{fig:NegControlRej}
    \vspace{-0.1in}
\end{figure}

Figure \ref{fig:NegControlRej} shows the percent of significant associations out of the 144 borders using the proposed resampling approach for the negative control analysis as a function of the smoothness parameter. The percentage of rejected tests for the naive test in Figure \ref{fig:negControlNaive} is far above the desired 0.05 level as we see roughly anywhere between \textcolor{black}{60\% to 80\%} rejection rates, with an increasing trend as a function of the buffer width. Given that this outcome should not be affected by police precincts, these results point to a lack of validity of the statistical test being run or the assumptions underlying the regression discontinuity design. With our resampling approach to inference, however, the results drop to a far more reasonable level with rejection rates close to 0.05 for each smoothness level.
Additionally, in Supplementary Materials Section 5, we show that the p-value histogram for each spatial smoothness value appears to be approximately uniformly distributed, as we would expect. Overall, the negative control analysis provides further justification for using our proposed resampling procedure, and gives increased belief in our findings on arrest rates in the following sections.

\subsection{Individual boundary estimates of precinct effects}
\label{subsec:NYCindividualresults}
Now that we have constructed null streets for each of the 144 precinct boundaries, we can perform hypothesis tests for each boundary to assess whether there is a causal effect of police precincts near the boundary between any two precincts. The bottom panel of Figure \ref{fig:NegControlRej} shows the results of these analyses by presenting the percentage of the 144 precinct boundaries for which the hypothesis test of no precinct effect was rejected using our proposed procedure. The naive hypothesis tests in Figure \ref{fig:NaivePvalue} show a large proportion of significant differences with more than 88\% of the tests being rejected. Using the proposed approach, this number is far smaller. The large difference in results between the two approaches highlights that either assumption 2b does not hold in this data set, or our statistical test is invalid. Nonetheless, the proportion of significant differences is still greater than 0.05 for most levels of spatial smoothness. For instance, at the \textcolor{black}{MSE-optimal} smoothing parameter (i.e., multiplier $= 1$), 14.4\% of the individual tests are rejected at the $\alpha = 0.05$ level, which is suggestive of a small amount of precinct-level differences in arrest rates. Note also that the percentage of rejections decreases as a function of the spatial smoothing multiplier, with a multiplier value of 4 leading to rejections in only 5.8\% of the precinct boundaries. Smoothing parameters that are 4 times the \textcolor{black}{MSE-optimal} choice lead to extreme oversmoothing, which tends to remove effects of finer-level spatial variability. While our tests should still be valid (in terms of type I error) in this setting, this may impact the power to detect effects. Although we include these results to illustrate our approach across a wide range of scenarios, we recommend focusing on smoothing parameter values near the \textcolor{black}{MSE-optimal} choice of $\sigma^2$, \textcolor{black}{especially since we acknowledge that there exists some sensitivity in type I error to the smoothness level of the surface.}


\subsection{Global variation in arrest rates}
\label{sec:GlobalNYC}

\begin{figure}[!htb]
\spacingset{1}
    \centering
        \includegraphics[width=0.5\linewidth]{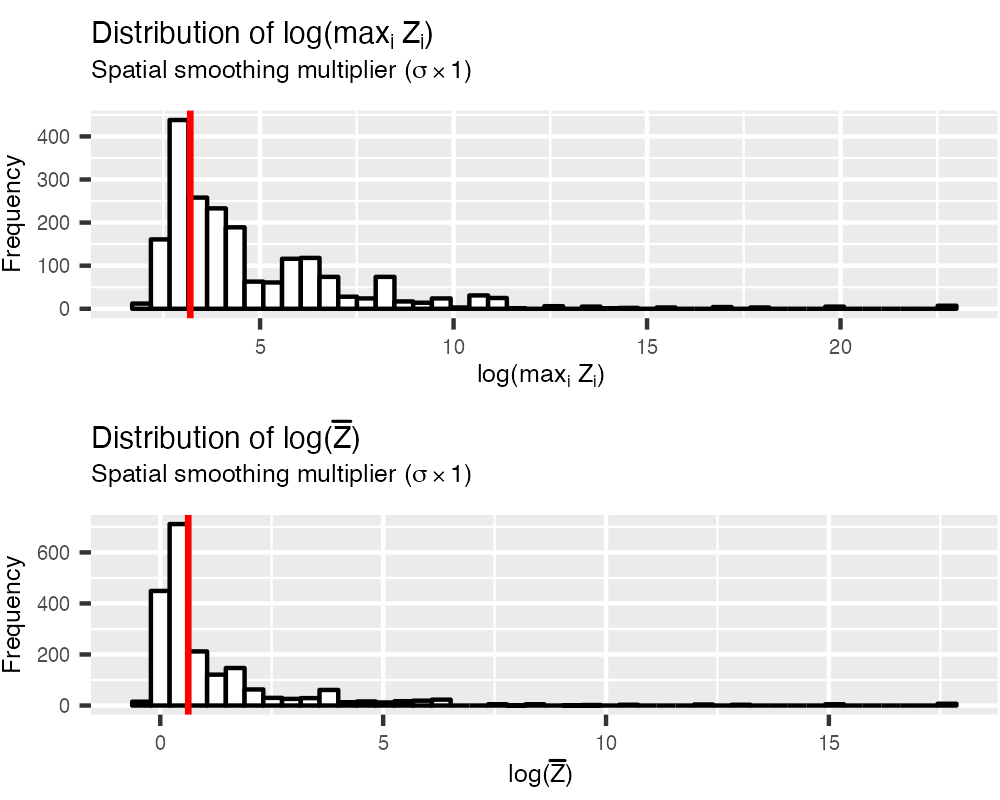}
        \caption{\textcolor{black}{Histograms of resampled test statistics from our proposed procedure using either $\max_i Z_i$ (top) or $\bar{Z}$ (bottom). The vertical lines are the observed test statistics computed at the actual precinct borders. All results are displayed on the log scale to improve visualization in the presence of outlying values.}}
    \label{fig:GlobalObsPval}
\end{figure}

In this section, we apply the approach of Section \ref{sec:Global} to assess whether there is an overall effect of police precincts on arrest rates across NYC. For comparison, we will calculate both $\bar{Z}$ and $\max_i Z_i$ as test statistics, defined in Section \ref{sec:Global}, to assess the magnitude of the overall precinct effect across the entire city.
To understand the distribution of each test statistic under the global null hypothesis of no precinct effects, we also calculate each test statistic using null streets to obtain $\bar{Z}^{(b)}$ and $\max_i Z_i^{(b)}$
for $b=1, \dots, B$, using the null streets discussed in \textcolor{black}{Section \ref{subsubsec:null}}. We perform this procedure $B$ times for each of the distinct levels of spatial smoothness. The results of this procedure can be seen in \textcolor{black}{Table \ref{tab:globalTab} and Figure \ref{fig:GlobalObsPval}}. We see that the p-value is quite large and above the $\alpha=0.05$ cutoff for all smoothness values \textcolor{black}{and for both test statistics}. As in Section \ref{subsec:NYCindividualresults}, the results tend to get more conservative as the smoothness parameter is increased. Figure \ref{fig:GlobalObsPval} shows the estimated null distribution of each test statistic for the \textcolor{black}{MSE-optimal} spatial smoothing value, and we see that the observed statistic (red vertical line) is well contained within the estimated null distribution in each case. This indicates that there is not a large degree of variation in policing practices across different precincts across NYC. While there may be differences at a small number of precincts, as indicated by the results in Section \ref{subsec:NYCindividualresults}, these effects appear to be relatively small and not widespread across the city.
\begin{table}[H]
    \centering
    \begin{tabular}{|c||l l l l l l|}
    \hline
    & \multicolumn{6}{c|}{Spatial smoothing multiplier}\\
     & 0.5 & 1.0 & 1.5 & 2.0 & 3.0 & 4.0\\
    \hline
    $\max_i Z_i$ & 0.340 & 0.670 & 0.662 & 0.838 & 0.878 & 0.891\\
    $\bar{Z}$ & 0.345 & 0.429 & 0.271 & 0.309 & 0.569 & 0.734\\
    \hline
    \end{tabular}
    \caption{\textcolor{black}{Results from the global test of variation in policing across NYC illustrating the p-value as a function of spatial smoothness.}}
    \label{tab:globalTab}
\end{table}

\section{Discussion}

In this manuscript, we first formalized estimands and developed estimation procedures for the geographic regression discontinuity design when the data follow a point process. Additionally, we proposed an approach to hypothesis testing for GeoRDDs that weakens the local randomization or continuity assumptions that are typically made in such studies. By leveraging the rich spatio-temporal information in our data on crime and arrests in NYC, we showed that valid hypothesis tests can be constructed even in the presence of certain violations of local randomization or continuity assumptions around the boundary of interest. The main idea is to find new boundaries that behave similarly to the boundary of interest, but are not near the border of two police precincts and therefore necessarily have no precinct effect. In the analysis of NYC arrest data, we found that analyses relying on a local randomization assumption lead to very strong conclusions that police precincts greatly impact arrest rates, while our approach based on resampling new streets leads to the conclusion that there is, at most, a small effect of police precincts on arresting practices. 

Our procedure was shown to work in a geographic regression discontinuity setting, though it is potentially applicable to other regression discontinuity settings as well. The only requirement is that new cutoffs of the running or forcing variable must be used where no treatment effect exists, and that the data set is rich enough to provide a large number of these new locations that are sufficiently independent of each other. While our procedure is able to provide statistical validity (type I error control) to tests of the hypothesis of no treatment effect, it is not able to correct for biases in estimation of treatment effects themselves. Further research is required to reduce the assumptions needed for estimation of treatment effects in regression discontinuity designs. \textcolor{black}{Additionally, future applications of our approach to areas outside of NYC with potentially different characteristics and population densities would offer an interesting new case study as well as insight into the robustness of our approach to different contexts.}

{
\spacingset{1}
\bibliographystyle{agsm}
\bibliography{References}
}

\newpage
{
\spacingset{1}
\section*{Supplementary Materials}
\asection{1}{Identification of causal effects}
\label{app:Identification}
Here we show how the causal effects of interest can be identified from the observed data. First, we examine $\theta(R_{\delta})$, which represents the average treatment effect within a distance of $\delta$ from the boundary. 
{\begin{align*}
\theta(R_{\delta}) &= E[Y^1(R_{\delta}) - Y^0(R_{\delta})] \\
&= E[Y^1(R_{\delta,1}) + Y^1(R_{\delta,0}) \\
& \quad \quad - Y^0(R_{\delta,1}) - Y^0(R_{\delta,0})] \\
&= E[Y^1(R_{\delta,1}) + Y^1(R_{\delta,1}) \\
& \quad \quad - Y^0(R_{\delta,0}) - Y^0(R_{\delta,0})] \quad \text{by assumption 2a} \\
&=2 E[Y^1(R_{\delta,1}) - Y^0(R_{\delta,0})] \\
&= 2 E[Y(R_{\delta,1}) - Y(R_{\delta,0})] \quad \text{by assumption 1}.
\end{align*}}
Next, we examine identification of $\tau(\boldsymbol{b})$ for any location $\boldsymbol{b} \in \mathcal{B}$. Identification of $\tau$ follows immediately as it is simply a weighted average of $\tau(\boldsymbol{b})$ for some weight function $w(\boldsymbol{b})$. Note here that we use the notation $\lim_{\boldsymbol{s} \to \boldsymbol{b}^1}$ to denote a limit that approaches the boundary location $\boldsymbol{b}$ from the precinct 1 side of the boundary, with an analogous notation for precinct 0. 
{\begin{align*}
    \tau(\boldsymbol{b}) &= \lambda^1(\boldsymbol{b}) - \lambda^0(\boldsymbol{b}) \\
    &= \lim_{\boldsymbol{s} \to \boldsymbol{b}^1} \lambda^1(\boldsymbol{b}) - \lim_{\boldsymbol{s} \to \boldsymbol{b}^0} \lambda^0(\boldsymbol{b}) \quad \text{by assumption 2b} \\
    &= \lim_{\boldsymbol{s} \to \boldsymbol{b}^1} \lambda(\boldsymbol{b}) - \lim_{\boldsymbol{s} \to \boldsymbol{b}^0} \lambda(\boldsymbol{b}) \quad \text{by assumption 1}.
\end{align*}}
Both of these terms in the final calculation are identifiable from the observed data as we can estimate the observed intensity surface on the two sides of the boundary, separately.

\asection{2}{Identification and estimation incorporating covariates}
\label{app:Covariates}

In this section, we show how relaxed identification assumptions that incorporate covariates can be used to identify $\tau(\boldsymbol{b})$. Throughout, it is assumed that we observe a vector of covariates at any spatial location $\boldsymbol{s}$, which we refer to as $\boldsymbol{V}(\boldsymbol{s})$. Note that this notation does not necessarily imply that these covariates are spatially correlated in the sense that $\text{cov}(\boldsymbol{V}(\boldsymbol{s}), \boldsymbol{V}(\boldsymbol{s}'))$ is higher when $\boldsymbol{s}$ and $\boldsymbol{s}'$ are closer to each other. It simply implies that the covariates have a distinct value at each location in the spatial domain. In the motivating study, these could represent variables such as socioeconomic status, which varies across the city. For this section, we must also define a spatial point process, which is a function of these covariates. For the observed data, we now have an intensity surface $\lambda(\boldsymbol{s}, \boldsymbol{v})$. This intensity surface is such that
$$E(Y(R)) = \int_{s \in R} \lambda(\boldsymbol{s}, \boldsymbol{V}(\boldsymbol{s})) d\boldsymbol{s},$$
for any region $R$. We have analogous definitions for the potential intensity surfaces under any particular treatment level $t \in \{0, 1\}$. In the context of $t=1$, we have $\lambda^1(\boldsymbol{s}, \boldsymbol{v})$, where the intensity surface is such that
$$E(Y^1(R)) = \int_{s \in R} \lambda^1(\boldsymbol{s}, \boldsymbol{V}(\boldsymbol{s})) d\boldsymbol{s}.$$
With these definitions in hand, we can proceed with identification of an analogous estimand to $\tau(\boldsymbol{b})$ that additionally incorporates covariates into the identification assumptions. 

Given that our estimand is itself defined in terms of potential intensity functions, which we have now defined to be functions of covariates, we must first adapt our estimand accordingly. Specifically, we focus on an estimand defined by
$$\tau(\boldsymbol{b}, \boldsymbol{V}(\boldsymbol{b})) = \lambda^1(\boldsymbol{b}, \boldsymbol{V}(\boldsymbol{b})) - \lambda^0(\boldsymbol{b}, \boldsymbol{V}(\boldsymbol{b})).$$
We have fixed the covariates at $\boldsymbol{V}(\boldsymbol{b})$, their value at the location of interest. We focus on this as our estimand because it most closely resembles $\tau(\boldsymbol{b})$ from the manuscript as it represents the treatment effect at the boundary of interest. Now that we have defined our modified estimand, we can also discuss modified identification assumptions that incorporate covariates as follows:

\textit{Assumption 2b incorporating covariates:} The potential outcome intensity surfaces satisfy
$$\lim_{\boldsymbol{s} \to \boldsymbol{b}} \lambda^t(\boldsymbol{s}, \boldsymbol{V}(\boldsymbol{b})) = \lambda^t(\boldsymbol{b}, \boldsymbol{V}(\boldsymbol{b})) \quad \text{for } t=0,1. $$
We can see how this assumption is weaker than Assumption 2b. If there is a discontinuity in the covariates at spatial location $\boldsymbol{b}$, then this will lead to a discontinuity in the potential intensity surface at $\boldsymbol{b}$ as well. This represents a violation in Assumption 2b, and we would incorrectly attribute this discontinuity to being a treatment effect if not addressed. This modified version of Assumption 2b allows there to be discontinuities in the covariates at $\boldsymbol{b}$ as long as the potential intensity surface is continuous with respect to $\boldsymbol{s}$ at $\boldsymbol{b}$ when the covariate values are set to $\boldsymbol{V}(\boldsymbol{b})$, their value at the location of interest. Under this assumption, we can write the estimand of interest as
\begin{align*}
    \tau(\boldsymbol{b}, \boldsymbol{V}(\boldsymbol{b})) &= \lambda^1(\boldsymbol{b}, \boldsymbol{V}(\boldsymbol{b})) - \lambda^0(\boldsymbol{b}, \boldsymbol{V}(\boldsymbol{b})) \\
    &= \lim_{\boldsymbol{s} \to \boldsymbol{b}^1} \lambda^1(\boldsymbol{s}, \boldsymbol{V}(\boldsymbol{b})) - \lim_{\boldsymbol{s} \to \boldsymbol{b}^0} \lambda^0(\boldsymbol{s}, \boldsymbol{V}(\boldsymbol{b})) \\
    &= \lim_{\boldsymbol{s} \to \boldsymbol{b}^1} \lambda(\boldsymbol{s}, \boldsymbol{V}(\boldsymbol{b})) - \lim_{\boldsymbol{s} \to \boldsymbol{b}^0} \lambda(\boldsymbol{s}, \boldsymbol{V}(\boldsymbol{b})).
\end{align*}
This is now a function of the observed data distribution that we can use the observed data to estimate. To estimate $\lim_{\boldsymbol{s} \to \boldsymbol{b}^1} \lambda(\boldsymbol{s}, \boldsymbol{V}(\boldsymbol{b}))$ we can use all data on the precinct 1 side of the boundary to estimate an intensity surface as a function of covariates. Standard software for estimating intensity functions that incorporate covariates available in the \texttt{R} package \texttt{spatstat} can be used. Additionally, observed data closer to the boundary $\boldsymbol{b}$ should receive more weight in this estimation process, and this is dictated by a spatial smoothness parameter, which can be estimated using cross-validation. While the main identification assumption is weakened somewhat by incorporating covariates, we must make an additional overlap assumption with respect to the covariates of interest. Let $H_1(\boldsymbol{v})$ be the density of $\boldsymbol{V}$ on the precinct 1 side of the boundary, and $H_0(\boldsymbol{v})$ be the density of $\boldsymbol{V}$ on the precinct 0 side of the boundary.\\
\\
\textit{Overlap assumption:} Both  $H_1(\boldsymbol{V}(\boldsymbol{b})) > 0$ and $H_0(\boldsymbol{V}(\boldsymbol{b})) > 0.$ \\
\\
This assumption states that covariate value $\boldsymbol{V}(\boldsymbol{b})$ must have positive density on both sides of the boundary. This is needed because we need to use data on both sides of the precinct separately to estimate the intensity surface at covariate value $\boldsymbol{V}(\boldsymbol{b})$. Without this assumption, we would be relying entirely on extrapolation to estimate this intensity function. 

Overall, these results show that if spatial covariates describing the area of interest are available, then they can be incorporated to weaken the identification assumptions that the GeoRDD relies upon in the point process setting. We focused on $\tau(\boldsymbol{b})$ throughout, but similar ideas could be used to incorporate covariates when identifying and estimating $\theta(R_{\delta})$. One could estimate an intensity surface that is a function of both $\boldsymbol{s}$ and $\boldsymbol{V}(\boldsymbol{s})$ using only data from the precinct 0 side of the boundary, and estimate what is expected to happen on the precinct 1 side of the boundary based on precinct 1's covariate values. A similar process would be done in the reverse order by using data from the precinct 1 side of the boundary to estimate what is expected to happen on the precinct 0 side of the boundary using precinct 0's covariate values. These would provide estimates of $Y^0(R_{\delta, 1})$ and $Y^1(R_{\delta, 0})$, which could be combined with the observed values for $Y^1(R_{\delta, 1})$ and $Y^0(R_{\delta, 0})$ in order to provide an estimate of $\theta(R_{\delta})$. This requires more extrapolation than for $\tau(\boldsymbol{b})$, however, as the intensity surface model must be extrapolated to a distance of $\delta$ from the boundary, as opposed to simply extrapolating to the border itself for $\tau(\boldsymbol{b})$.

\asection{3}{Theoretical derivations for estimating the null distribution}
\label{app:A}

Here we provide the full mathematical details of the results shown in Section 3.5. First, we can show the type I error rate can be written as:
{\begin{align*}
    P(\text{reject } H_0 \; | \; H_0) &= \int_q P(\text{reject } H_0 \; | \; H_0; \hat{Q}_{1-\alpha} = q) \cdot f_{\hat Q} (q) dq\\
    &= \int_q P(Z_i > q; X_i) \cdot f_{\hat Q} (q) dq\\
    &= \int_q \qty[1- P(Z_i \leq q; X_i) ] \cdot f_{\hat Q} (q) dq\\
    &= 1 - \int_q P(Z_i \leq q; X_i) \cdot f_{\hat Q} (q) dq\\
    &= 1 - E_{\hat Q} \qty[F(X_i, q)].
\end{align*}}
This shows that we need $E_{\widehat Q} \qty[F(X_i, \widehat{Q}_{1-\alpha})] \geq 1- \alpha$ in order to obtain type I error control. Next we highlight properties of our estimate of the CDF of the null distribution of the test statistic. First, we can show that the mean of this estimate can be written as:
{\begin{align*}
    E[\hat{F}(X_i, q)] &= E_{X^{(b)}}\qty[E\qty[\hat{F}(X_i, q) \Big | X_{i}^{(1)}, X_{i}^{(2)}, ..., X_i^{(B)}]] \\
    &= E_{X^{(b)}}\qty[\frac{1}{B}\sum_{b=1}^B P(Z_i^{(b)} \leq q; X_i^{(b)})]\\
    &= E_{X^{(b)}}\qty[\frac{1}{B}\sum_{b=1}^B F(X_i^{(b)}, q)]\\
    &= E_{X^{(b)}}\qty[F(X_i^{(b)}, q)]\\
    &\approx F(X_i, q) + \frac{d}{dX} F(X_i, q) \cdot E(X_i^{(b)} - X_i)\\
    & \quad \quad + \frac{d^2}{dX^2} F(X_i, q) \cdot E\qty[(X_i^{(b)} - X_i)^2]
\end{align*}}
This shows that the mean of the CDF estimate depends on how far off the covariates in the null streets, $X_i^{(b)}$, are from the covariates at the precinct boundary of interest, denoted by $X_i$. Lastly, we can write the variance of our CDF estimate as:
{\begin{align*}
    Var[\hat{F}(X_i, q)] &= E\qty[Var\qty[\hat{F}(X_i, q) \Big | X^{(b)}]] + Var\qty[E\qty[\hat{F}(X_i, q) \Big | X^{(b)}]] \\
    &= E\qty[ \frac{1}{B^2} \sum_{b=1}^B F(X_i^{(b)}, q) \cdot (1-F(X_i^{(b)}, q))] + Var\qty[\frac{1}{B} \sum_{b=1}^B F(X_i^{(b)}, q)]\\
    &= \frac{1}{B} E\qty[ F(X_i^{(b)}, q) \cdot (1-F(X_i^{(b)}, q))]+ \frac{1}{B} Var[F(X_i^{(b)}, q)]\\
    &\approx \frac{1}{B} \Bigg[E\qty[F(X_i^{(b)}, q) \cdot (1-F(X_i^{(b)}, q))] + Var\qty[F(X_i; q) + \frac{d}{dX_i} F(X_i, q) (X_i^{(b)} - X_i)]\Bigg]\\
    &= \frac{1}{B} \Bigg[E\qty(F(X_i^{(b)}, q) \cdot (1-F(X_i^{(b)}, q))) + \qty(\frac{d}{dX_i}F(X_i, q))^2 \cdot Var(X_i^{(b)})\Bigg]
\end{align*}}
For simplicity of exposition, all of these results utilized a scalar covariate $X_i$, but could be easily extended to accommodate a vector of covariates to match on and analogous results would hold. 

\asection{4}{Investigating covariates used to select null streets}
\label{app:B}
As illustrated in the manuscript, it is important that we sample null streets with similar values of important covariates as the precinct boundaries of interest. This will help to ensure the test statistics at the null streets have a similar distribution as at the precinct borders under the null hypothesis. \textcolor{black}{Figures \ref{fig:ex1}, \ref{fig:ex2}, and \ref{fig:ex3} show the mean and variance of the test statistics at all possible null streets after being binned according to the $t^{\text{sum}}(\cdot)$ and $t^{\text{ratio}}(\cdot)$ values of the respective covariate used in the arrest data analysis, negative control, and simulation, respectively. Note that the log of the test statistics are used to improve visualization in the presence of outlying values. In Figure \ref{fig:ex1}, we see a clear relationship between the ratio and total crime with the mean and variance of the test statistics. Figure \ref{fig:ex2} illustrates similar results for the relationship between the ratio and total street length with the mean and variance of the test statistics for the negative control. Lastly, Figure \ref{fig:ex3} illustrates the relationship between the ratio and total buffer area with the mean and variance of the test statistics for the simulation. This motivates our procedure for finding null streets based on these variables in the different analyses.}

\begin{figure}[H]
    \centering
    \includegraphics[width=0.7\linewidth]{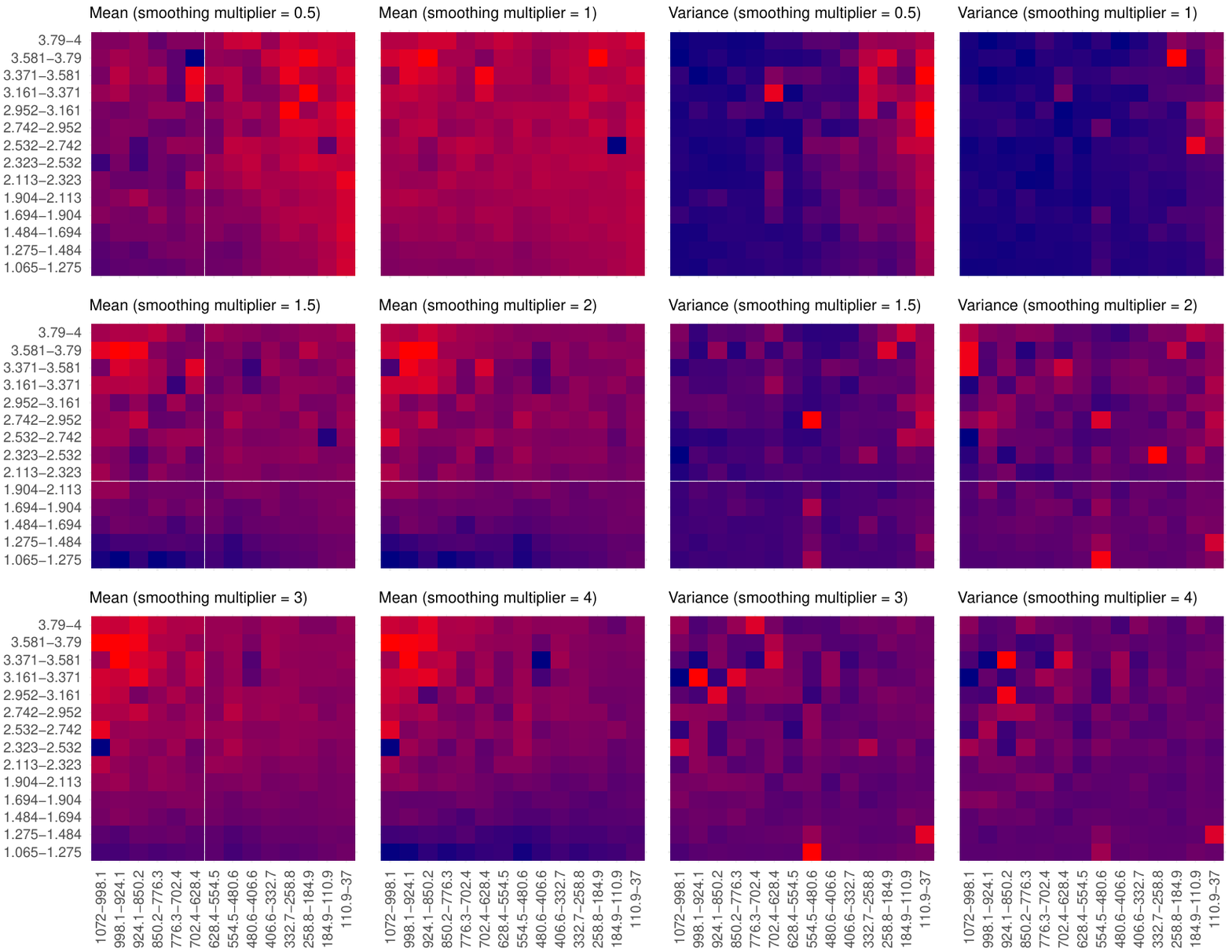}
    \caption{Heat map illustrating the mean (left half) and variance (right half) of the test statistics (on log scale) for the arrest data analysis at the null streets as a function of the ratio of crime (y-axis) and the total amount of crime (x-axis) for varying degrees of spatial smoothness.}
    \label{fig:ex1}
\end{figure}

\begin{figure}[H]
    \centering
    \includegraphics[width=0.7\linewidth]{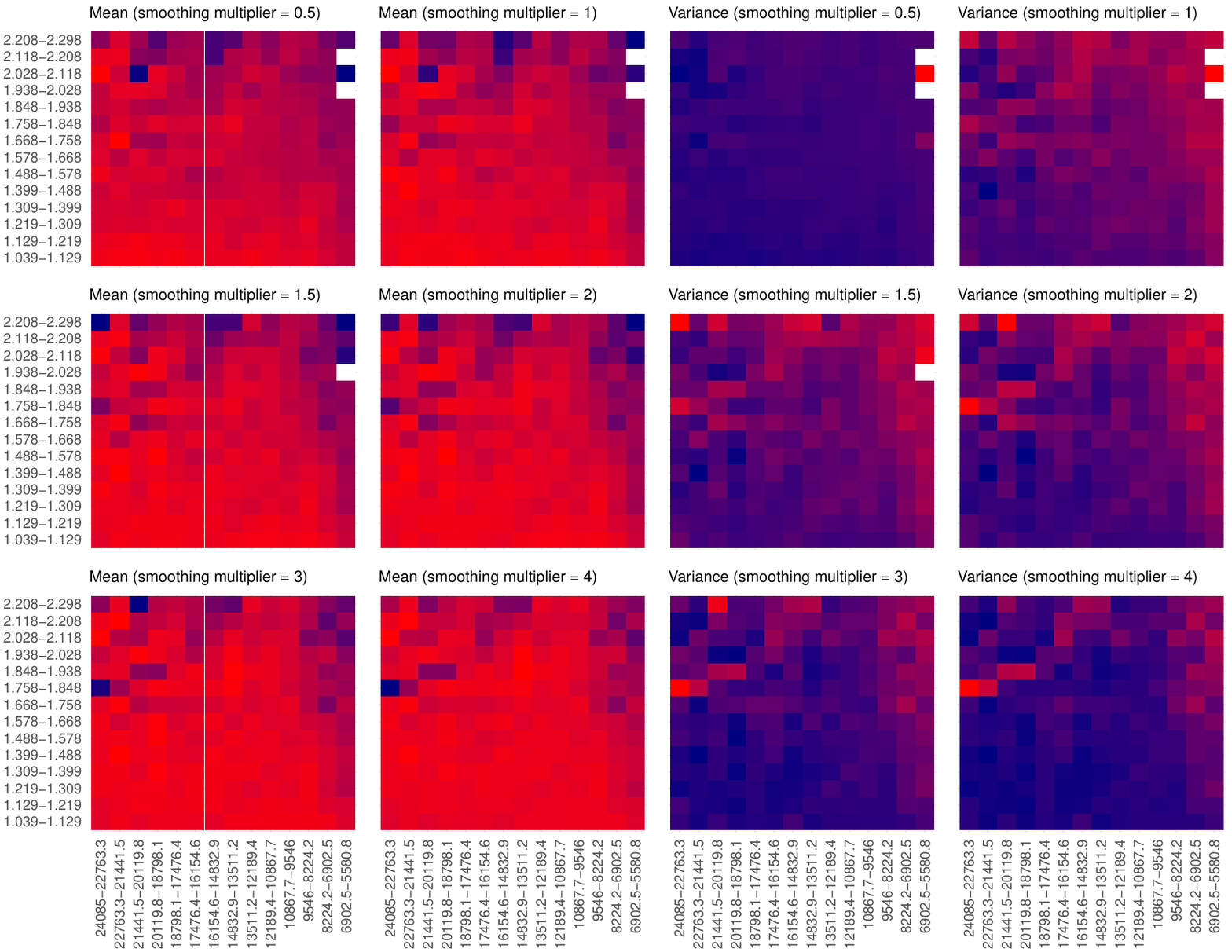}
    \caption{Heat map illustrating the mean (left half) and variance (right half) of the test statistics (on log scale) for the negative control at the null streets as a function of the ratio of street lengths (y-axis) and the total street length (x-axis) for varying degrees of spatial smoothness.}
    \label{fig:ex2}
\end{figure}

\begin{figure}[H]
    \centering
    \includegraphics[width=0.7\linewidth]{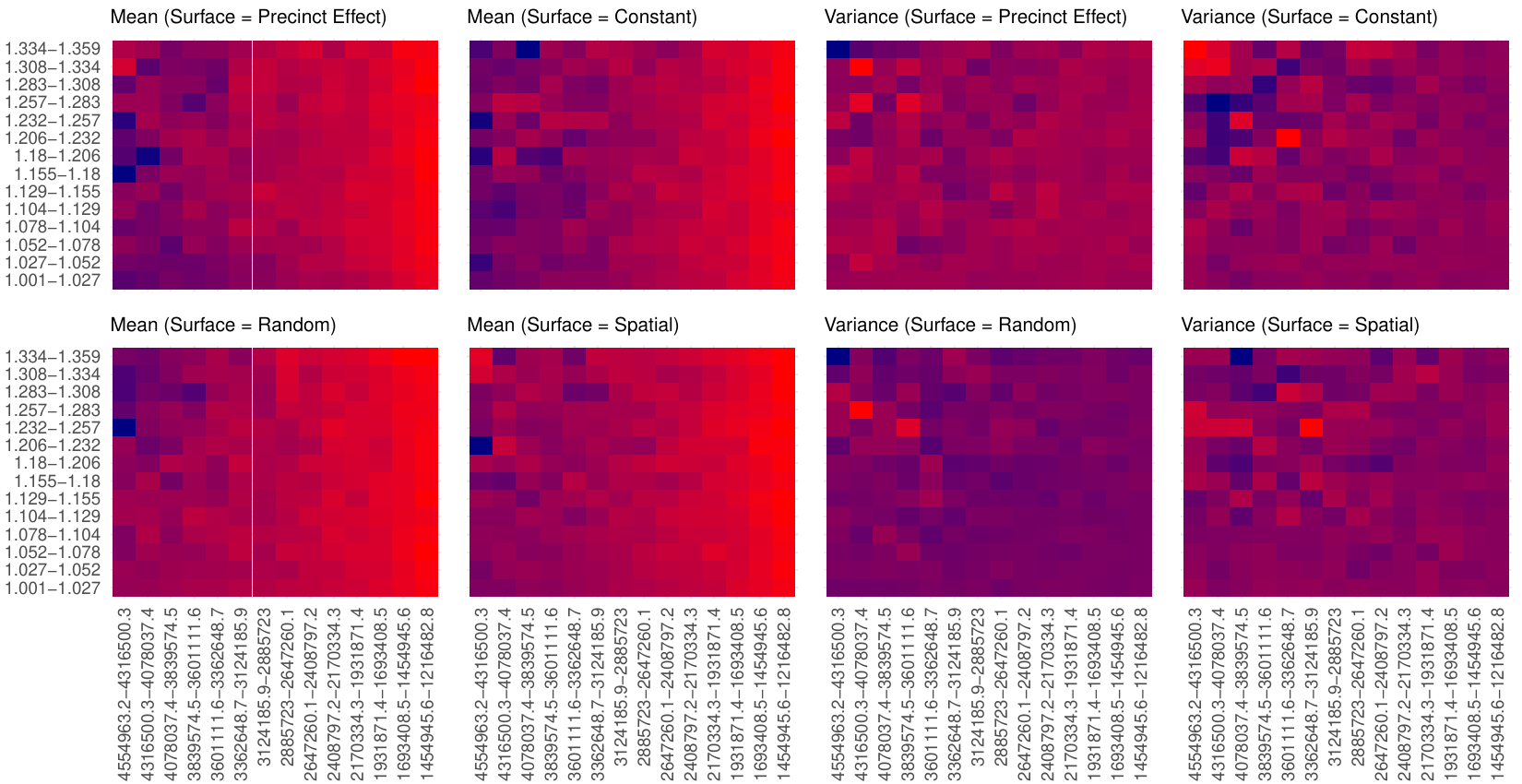}
    \caption{Heat map illustrating the mean (left half) and variance (right half) of the test statistics (on log scale) for the simulation at the null streets as a function of the ratio of buffer areas (y-axis) and the total buffer area (x-axis) for the different simulated surfaces.}
    \label{fig:ex3}
\end{figure}

\asection{5}{Additional results using $\vb*{\tau}$ as estimand}
In this section, we present additional figures and results for the NYC policing analysis and the negative control analysis when using $\tau$ as an estimand. First, we show the global test results in the negative control analysis, where we see large p-values regardless of the global test statistic being used or the spatial smoothness level examined. This is expected given that this outcome is a negative control and should not be affected by police precincts. We then provide histograms of p-values from the individual precinct boundary tests across all 144 precincts. We do this for both the negative control analysis and the arrest analysis for a wide range of spatial smoothing parameters. We see that the histograms for the arrest data have slightly more values closer to zero than what one would expect under the null hypothesis of no precinct effect, which highlights the findings in Section 5.3 of the manuscript where more than 5\% of the hypothesis tests were rejected. This decreases somewhat as we increase the smoothness of the intensity surface estimates. For the negative control, the histograms appear relatively uniformly distributed, which is again expected given that these outcomes should not be affected by police precincts. These further highlight the ability of our approach to provide valid hypothesis tests in this setting. Lastly, we provide additional results about the choice of $c$ when determining null streets for the negative control analysis. In the manuscript, we showed how results were sensitive to this choice, and that values closer to 1 gave the desired type I error rate. Here we show the same plots across all smoothness parameter values, and find largely the same results, showing the importance of $c$ in the process for choosing null streets. 

\subsection{Global test results for negative control}
\begin{center}
    \includegraphics[width=0.495\linewidth]{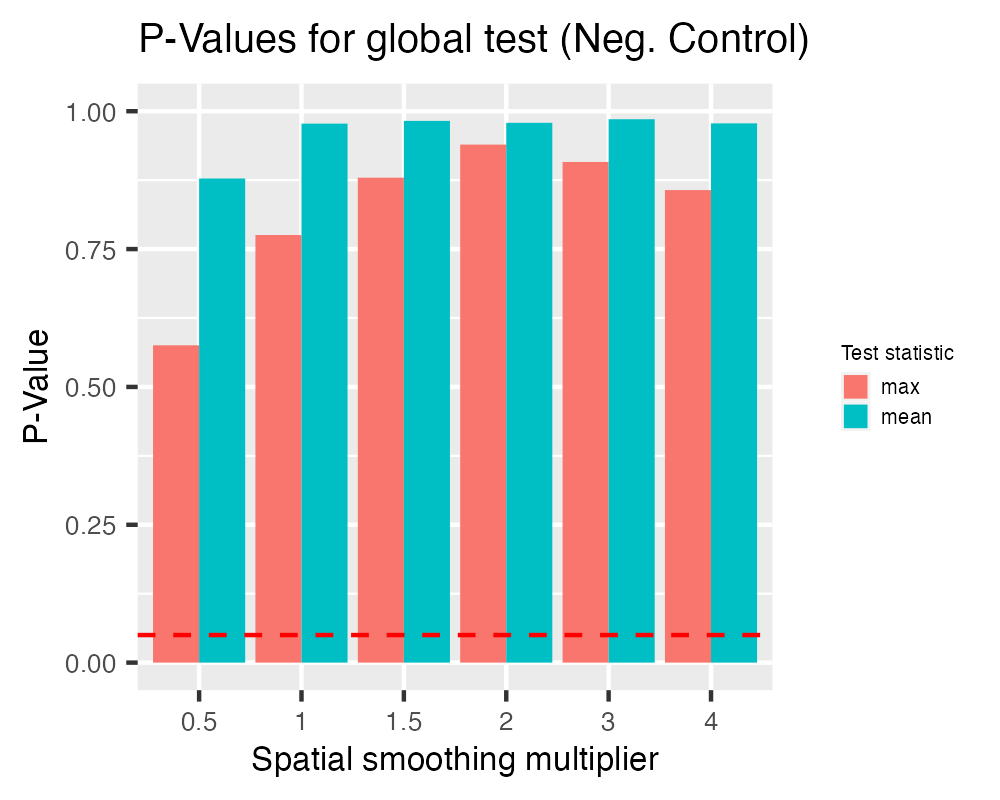}  
\end{center}

\subsection{Histograms of the corrected p-values for various spatial smoothing levels}
\begin{center}
    \includegraphics[width=\linewidth]{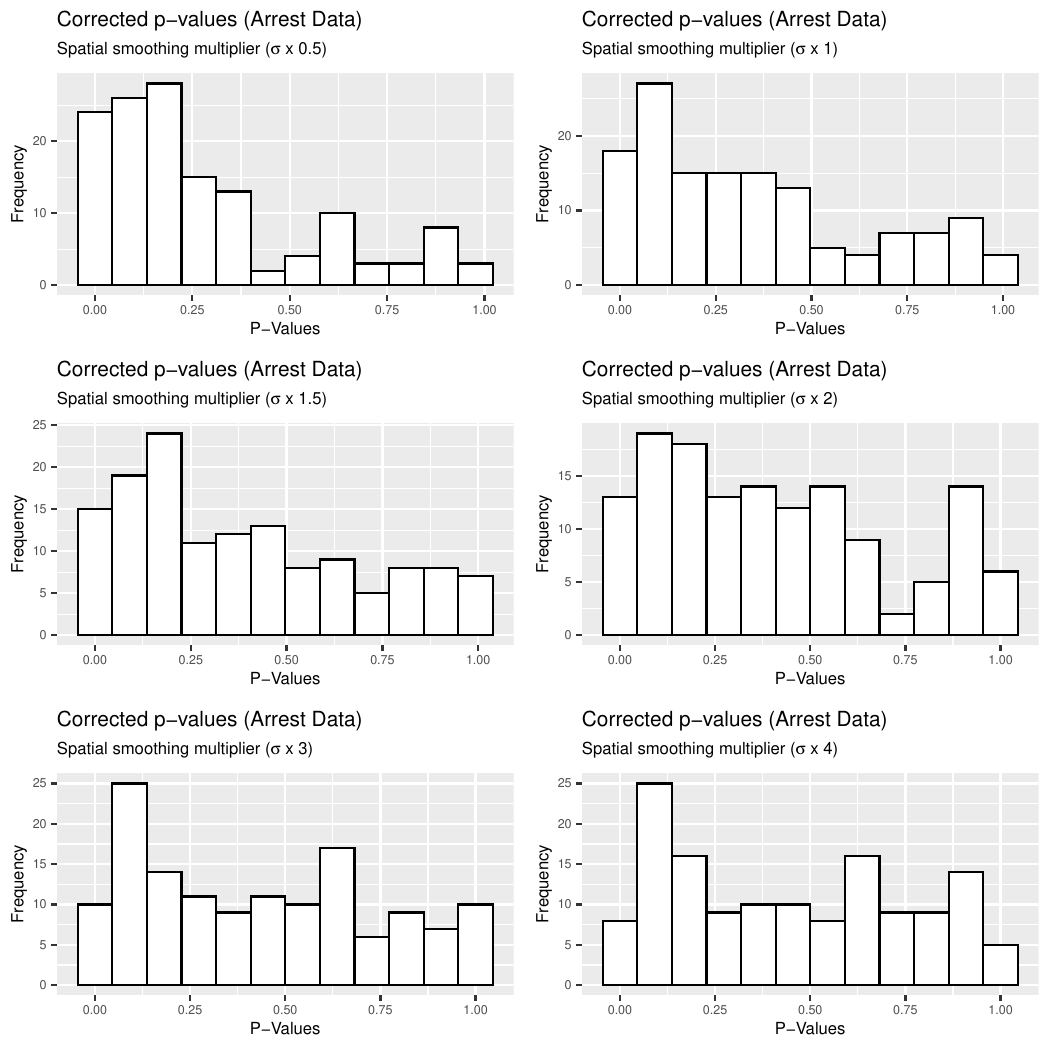}
\end{center}
\begin{center}
    \includegraphics[width=\linewidth]{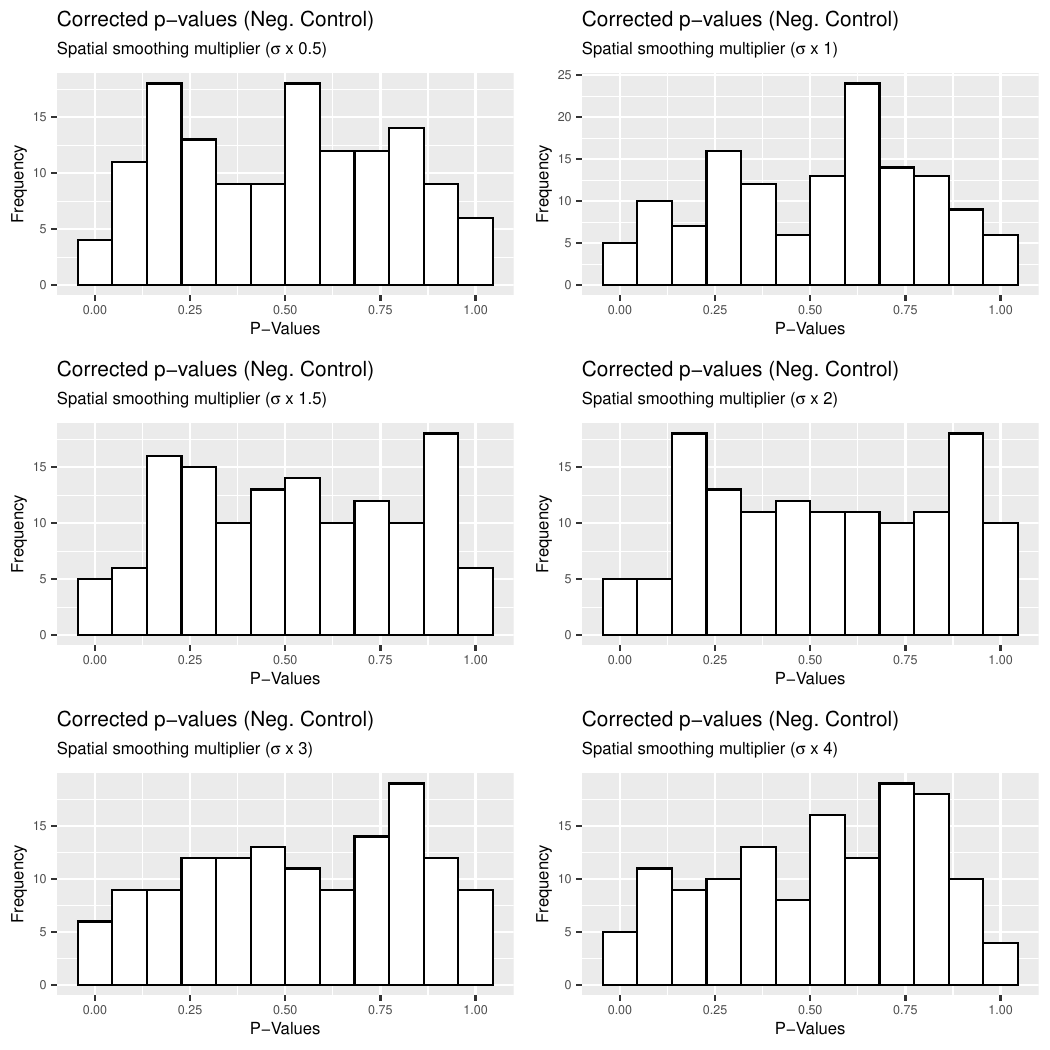}
\end{center}

\subsection{Type I error as a function of $c$}
\begin{center}
    \includegraphics[page = 1, width = 0.49\linewidth]{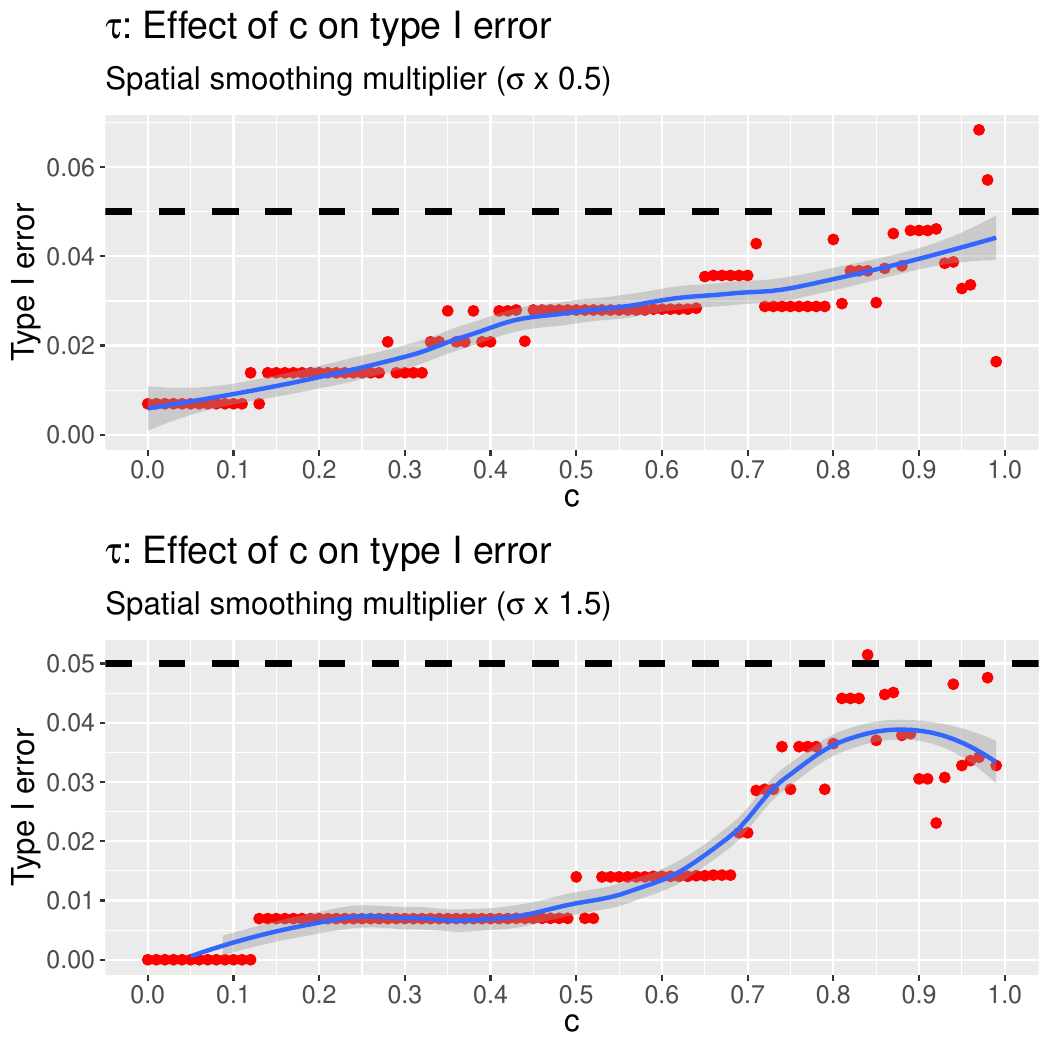}
    \includegraphics[page = 2, width = 0.49\linewidth]{supp_plots/match_tau.pdf}
    \includegraphics[page = 3, width = 0.49\linewidth]{supp_plots/match_tau.pdf}
    \includegraphics[page = 4, width = 0.49\linewidth]{supp_plots/match_tau.pdf}
\end{center}

\asection{6}{Results using $\vb*{\theta(R_\delta)}$ as estimand}
Here we show the same results as in the manuscript, where instead we use $\theta(R_\delta)$ as the estimand instead of $\tau$. We show results for both the arrest outcome as well as the negative control outcome for a variety of buffer widths, denoted by $\delta$. Note throughout this section that our outcomes $Y(R^{(i)}_{\delta,0})$ and $Y(R^{(i)}_{\delta,1})$ are the counts of the number of events within a buffer width of $\delta$ around boundary $i$, divided by a scaling factor dependent on the outcome being examined. For arrests, we divide the total number of arrests by the amount of crime in a region to examine arrest rates instead of counts. For the negative control outcome, we scale the number of trees by the total length of streets in the region to obtain a rate of trees per length of street. 

We first show the percentage of rejections across the 144 police precinct boundaries for both arrests and the negative control outcome. As expected, for the negative control outcome we see rejections in approximately 5\% of the tests across all buffer widths, highlighting the utility of our proposed approach to inference. For the arrest data, we see similar results as for the $\tau$ estimand, though they are slightly more conservative. At small buffer widths, we see more than the 5\% of rejections we would expect if there were no precinct effects, suggesting a small effect of precincts across the city. This effect dissipates, however, as the buffer width increases. We next show the results of the global test for both the mean and max as the global test statistic. For the negative control outcome, all of the p-values are well above 0.05 as expected given that this outcome should not be affected by police precincts. For the arrest data, at the smallest buffer width of 300 feet the p-values are smaller, yet still above the 0.05 threshold. These increase as the buffer width grows, which mirrors the results above for the individual tests showing some significant tests at a buffer width of 300 or 400 feet, which then disappear at larger buffer widths. We also show the p-value histograms, which mirror the results seen for $\tau$. For the negative control outcome these are largely uniformly distributed, while for the arrest data they assign more weight to small p-values at lower buffer widths. Lastly, we investigate the choice of $c$ on the type I error in the negative control outcome for $\theta(R_{\delta})$. As for the $\tau$ estimand, the choice of $c$ is very important for type I error control, though the value of $c$ needed for valid type I error control is different for this estimand. For this estimand, small values of $c$ also lead to conservative inference, but larger values of $c$ lead to anti-conservative inference. The optimal choice of $c$ depends slightly on the smoothing parameter, but a value of $c$ around 0.65 leads to good type I error control in all analyses, which is therefore the value that we proceed with.

\subsection{Proportion of p-values less than 0.05 using the corrected testing approach}
\begin{center}
    \includegraphics[width=0.495\linewidth]{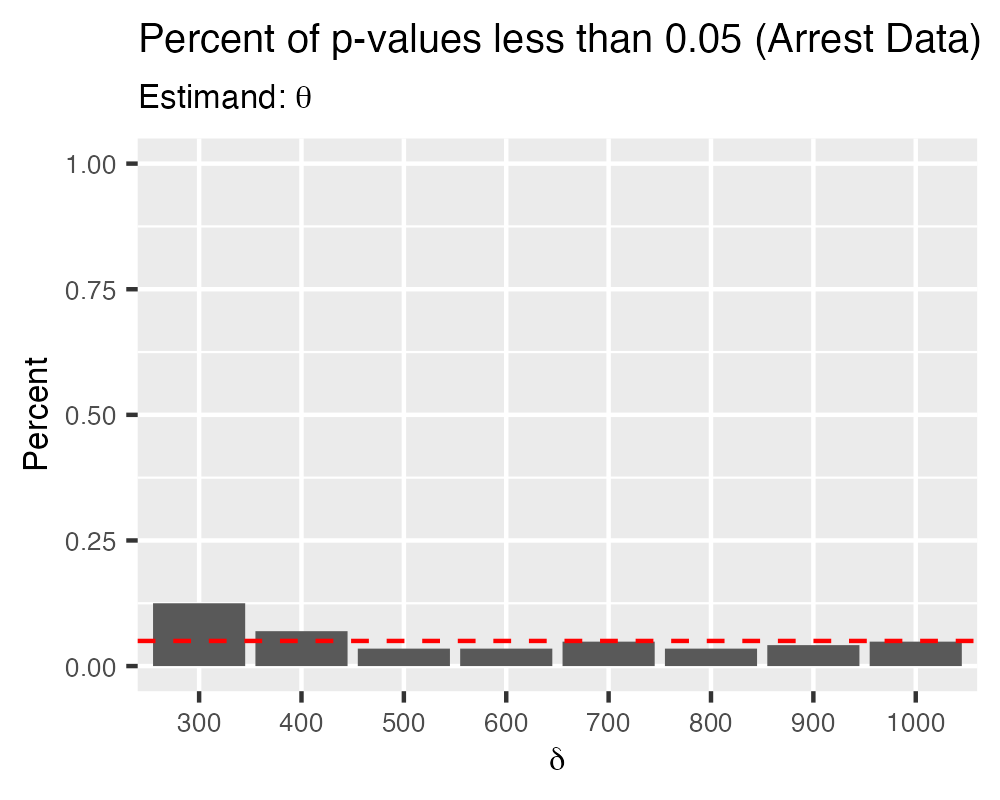}
    \includegraphics[width=0.495\linewidth]{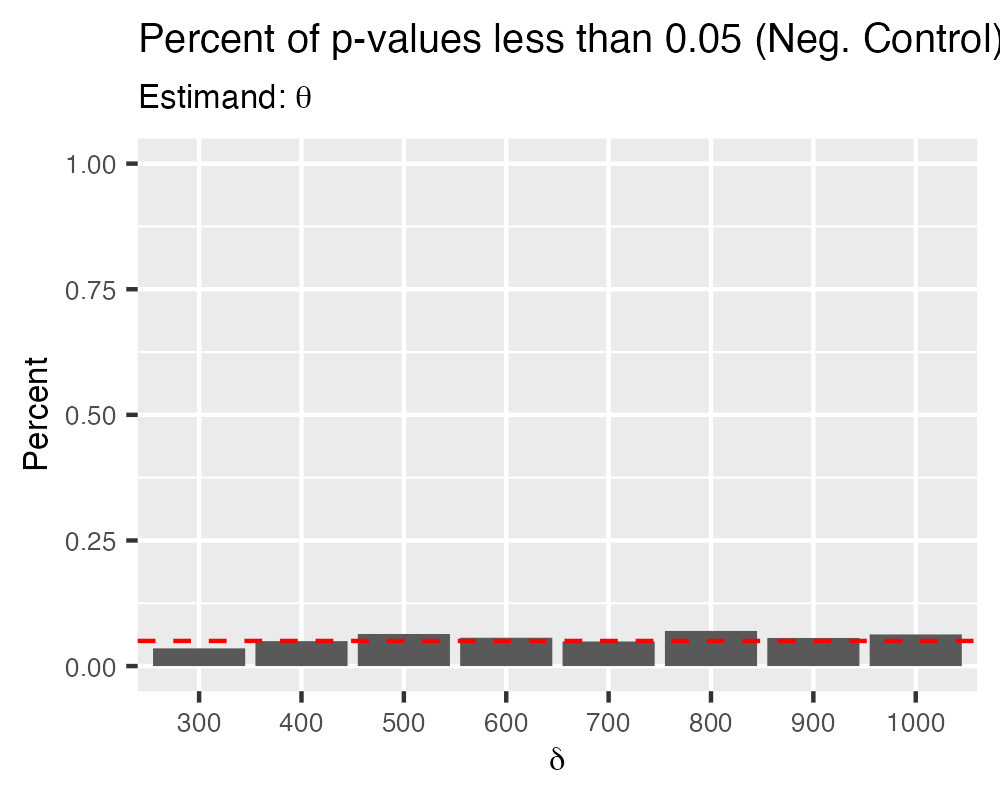}
\end{center}

\subsection{Global test results}
\begin{center}
    \includegraphics[width=0.495\linewidth]{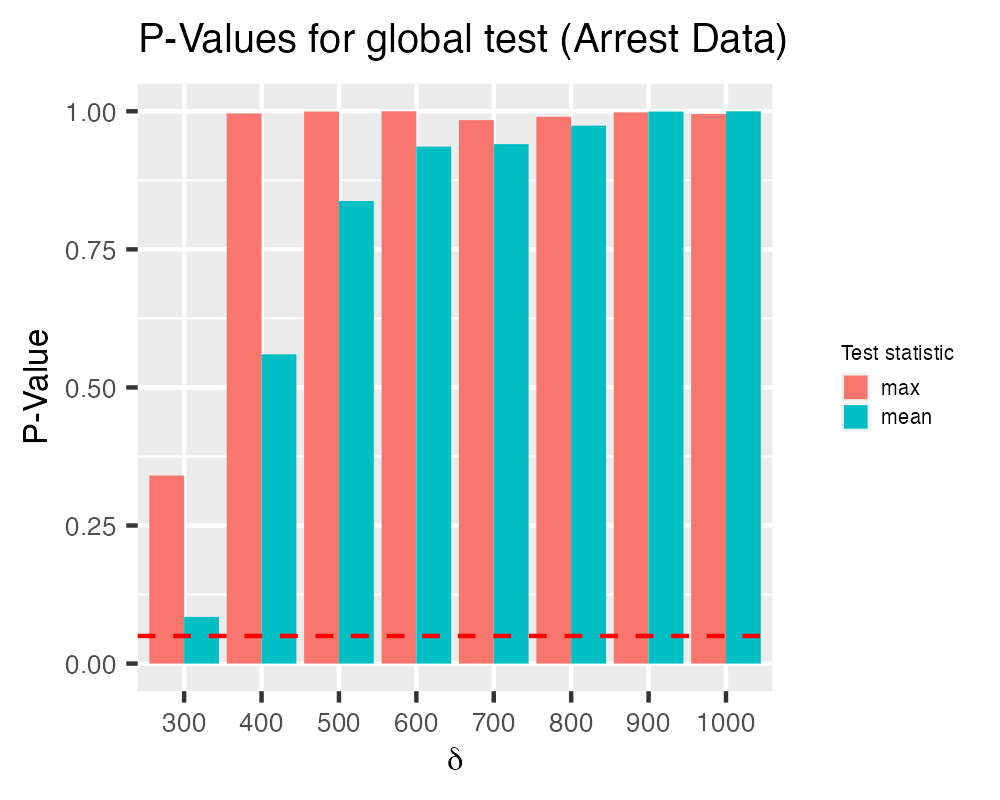}   
    \includegraphics[width=0.495\linewidth]{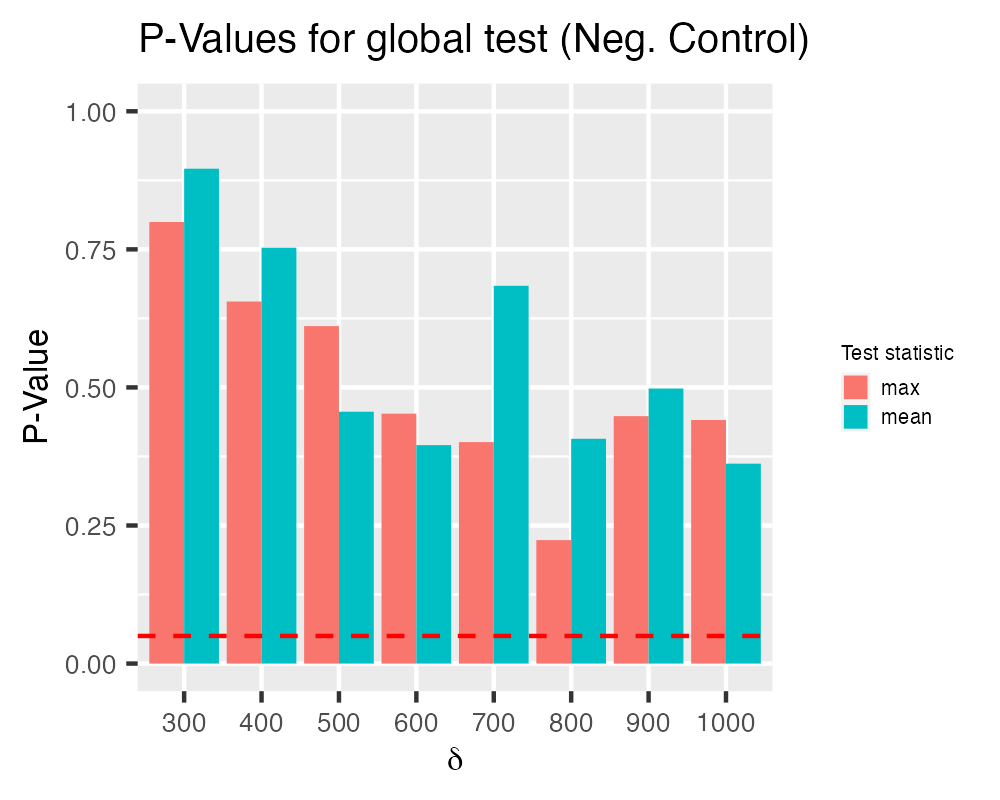} 
\end{center}

\subsection{Histograms of the corrected p-values for various $\delta$}
\begin{center}
    \includegraphics[width=\linewidth]{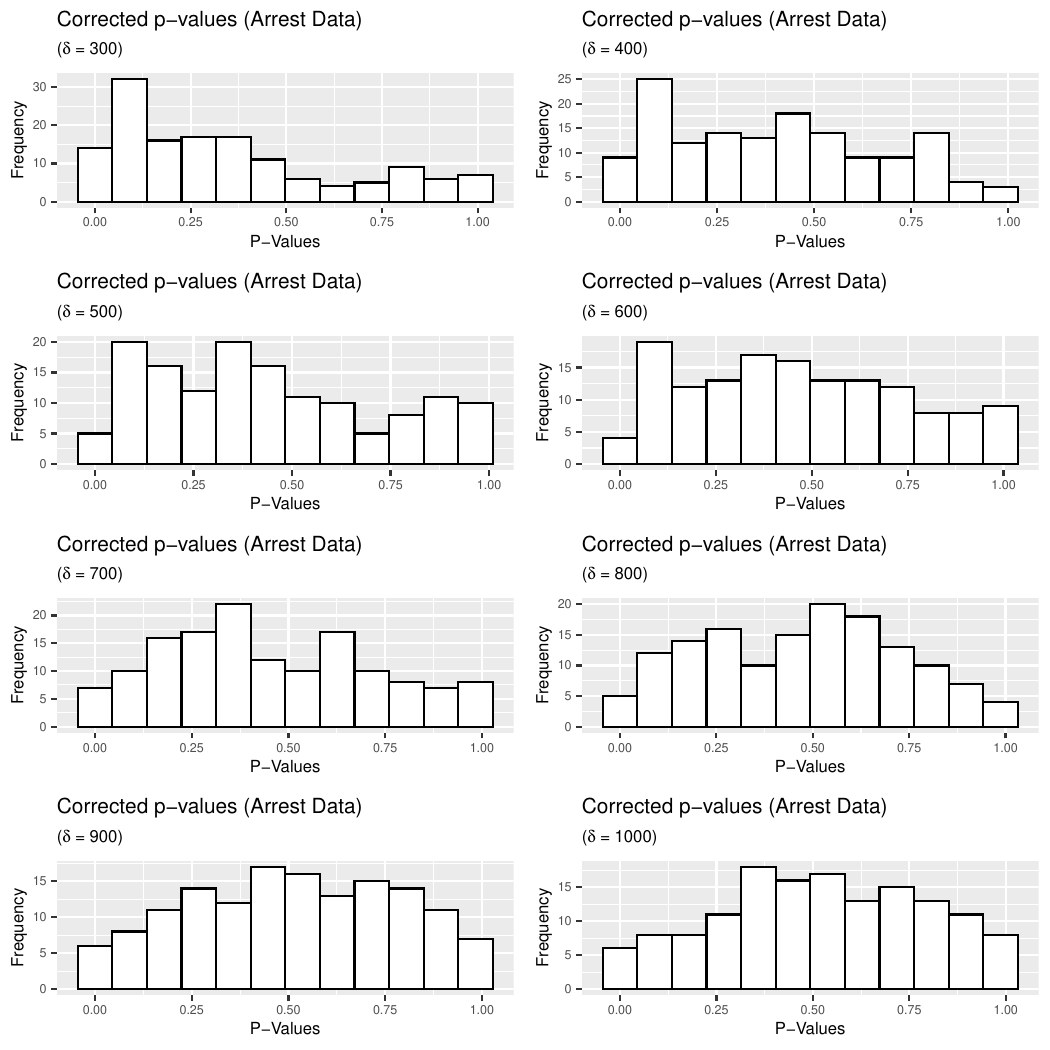}
\end{center}
\begin{center}
    \includegraphics[width=\linewidth]{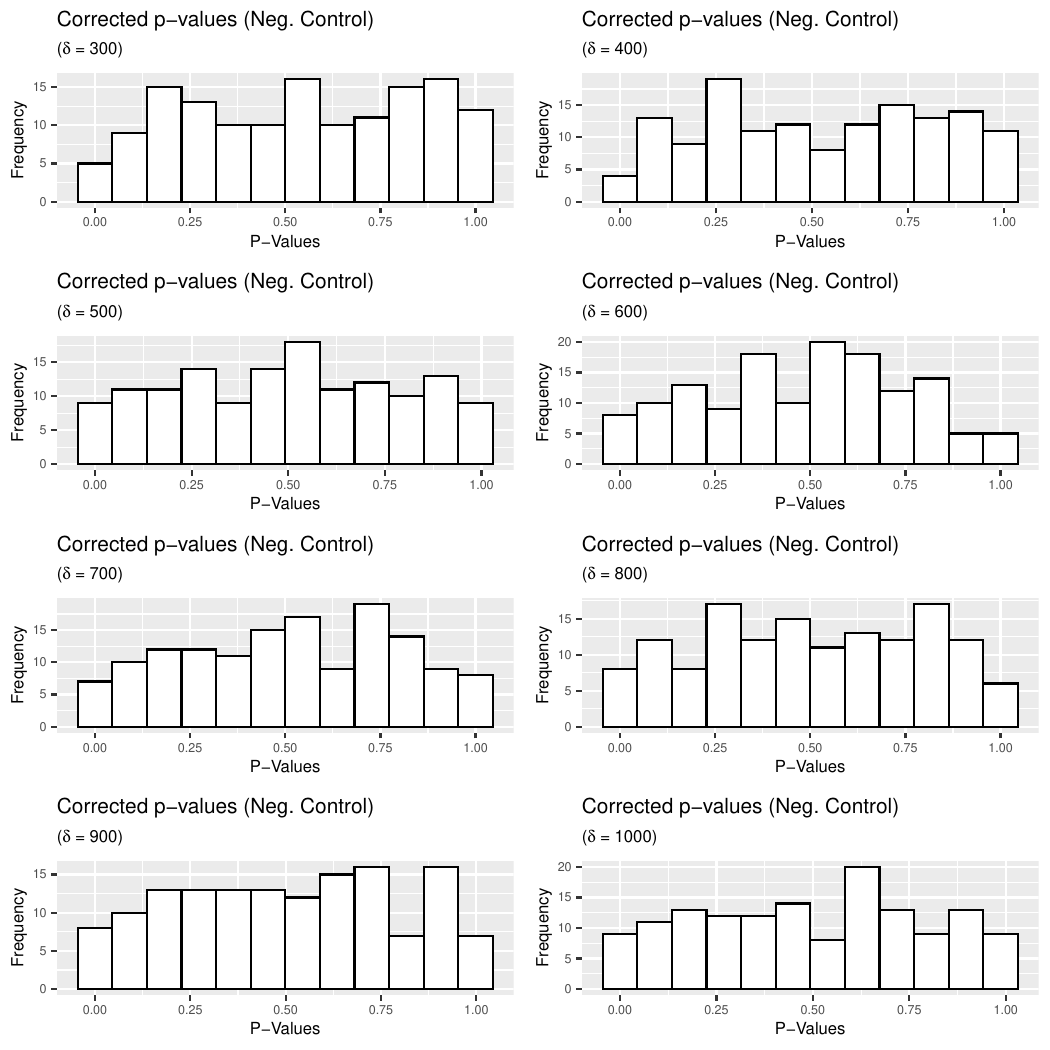}
\end{center}

\subsection{Type I error as a function of $c$}
\begin{center}
    \includegraphics[page = 1, width = 0.49\linewidth]{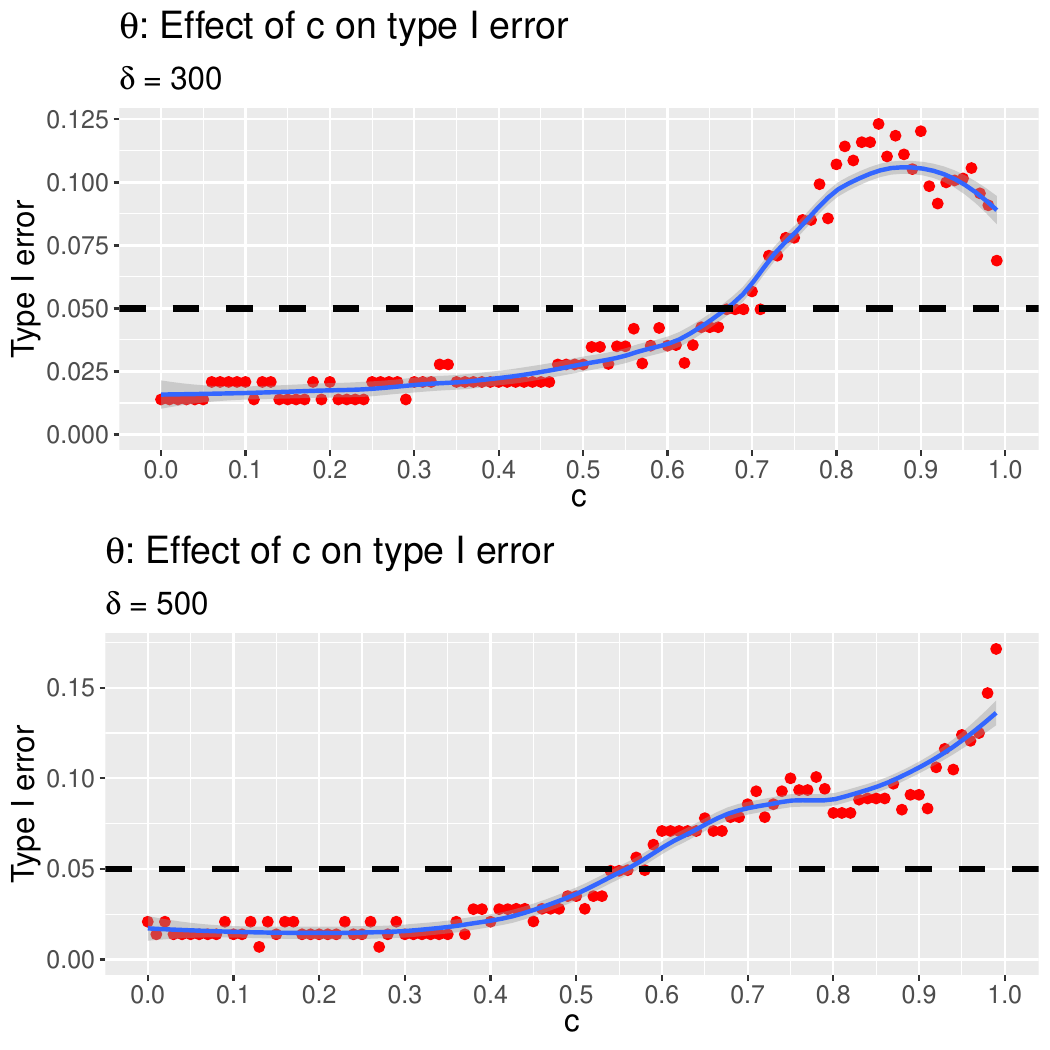}
    \includegraphics[page = 2, width = 0.49\linewidth]{supp_plots/match_theta.pdf}
    \includegraphics[page = 3, width = 0.49\linewidth]{supp_plots/match_theta.pdf}
    \includegraphics[page = 4, width = 0.49\linewidth]{supp_plots/match_theta.pdf}
\end{center}

\asection{7}{Map of all NYC precincts and streets}
Here we show a map of all streets available in the NYC analysis. The red lines correspond to all streets in NYC that are available to be null streets, while the black lines correspond to precinct boundaries. 
\begin{center}
    \includegraphics[scale = 0.5,trim={20cm 3cm 20cm 0cm},clip]{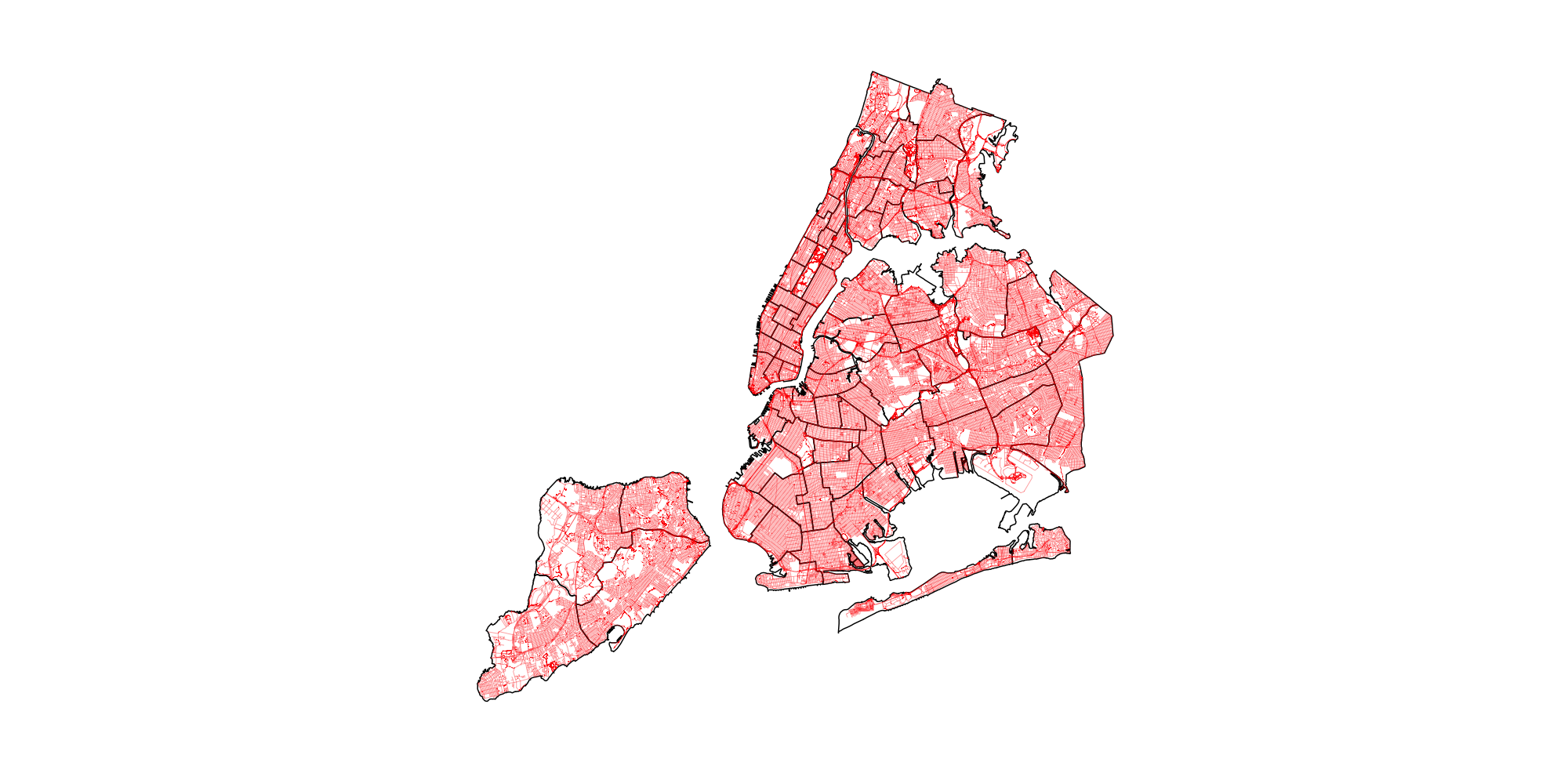}
\end{center}

\asection{8}{Simulation power analysis}
\textcolor{black}{In order to test the power of our approach under differing sample sizes, we have performed an additional simulation study, but with smaller sample sizes. Our simulated data are simulated ``counts'' of observations in specified regions. These data are produced by generating a Poisson random variable with mean equal to the integrated intensity surface over the region of interest. Therefore, the notion of a changing ``sample size'' is addressed by changing the mean of the Poisson point process. In order to simulate a smaller sample size, we can simply reduce the magnitude of the intensity surface over all of NYC which in turn will reduce the simulated observation counts. Hence, for all four of our different simulated surface types, we have reduced the magnitude of the intensity surface by a factor of two over all of NYC and then run our analyses again. As shown in Table \ref{tab:simP2}, our procedure produces very similar results to those from the manuscript, with only slight decreases in power for the individual level tests. Importantly, type I error remains relatively unchanged across the four scenarios showing the validity of our approach.}
\begin{table}[!htb]
    \centering
    \begin{tabular}{|l||l|l|l|l| }
    \hline
    \multicolumn{5}{|c|}{Individual}\\
    \hline
    $\delta$ &Constant &Random &Spatial &Precinct \\
    \hline
    300 & 0.050 & 0.052 & 0.050 & 0.904\\
    400 & 0.050 & 0.052 & 0.053 & 0.921\\
    500 & 0.050 & 0.051 & 0.053 & 0.929\\
    600 & 0.050 & 0.052 & 0.059 & 0.941\\
    700 & 0.051 & 0.054 & 0.066 & 0.95\\
    800 & 0.053 & 0.056 & 0.071 & 0.955\\
    900 & 0.051 & 0.055 & 0.074 & 0.959\\
    1000 & 0.053 & 0.057 & 0.079 & 0.963\\
    \hline
    \hline
    \multicolumn{5}{|c|}{Global ($\max_i Z_i$; $\bar{Z}$)}\\
    \hline
    $\delta$ &Constant &Random &Spatial &Precinct \\
    \hline
    300 & 0.049; 0.049 & 0.028; 0.035 & 0.034; 0.036 & 1.000; 1.000\\
    400 & 0.034; 0.043 & 0.031; 0.038 & 0.028; 0.041 & 1.000; 1.000\\
    500 & 0.051; 0.032 & 0.026; 0.023 & 0.032; 0.048 & 1.000; 1.000\\
    600 & 0.031; 0.027 & 0.038; 0.036 & 0.047; 0.067 & 1.000; 1.000\\
    700 & 0.049; 0.041 & 0.041; 0.057 & 0.058; 0.156 & 1.000; 1.000\\
    800 & 0.041; 0.040 & 0.038; 0.064 & 0.060; 0.224 & 1.000; 1.000\\
    900 & 0.067; 0.050 & 0.038; 0.056 & 0.083; 0.253 & 1.000; 1.000\\
    1000 & 0.034; 0.053 & 0.041; 0.058 & 0.088; 0.345 & 1.000; 1.000\\
    \hline
    \end{tabular}
    \vspace{0.1in}
    \caption{Probability of rejecting the null hypothesis across the four simulation scenarios and differing buffer widths. The top half of the table corresponds to hypothesis tests at individual precinct boundaries, while the bottom half corresponds to the global test of variation across NYC using two different test statistics.}
    \label{tab:simP2}
    \vspace{-0.1in}
\end{table}

\asection{9}{Residential impact: precincts and other sub-municipal geographies}
\textcolor{black}{The key assumptions required for the GeoRDD in the context of NYC police precincts would likely be violated if police precincts systematically influenced where people decide to reside. To better understand this issue, we seek to uncover whether people might be choosing where to live based on the police precinct. Hence, we investigated, first, how NYC residents decide where to live, and, second, whether police precinct boundaries overlap with any other important government boundaries.}
    
\textcolor{black}{On the first point, \textit{Upsold} is a sociological study ``about consumers’ preferences and decision-making in the context of purchasing homes'' \citep[][9]{besbris2020upsold}. In following fifty-seven NYC home buyers through the process of their purchase, \cite{besbris2020upsold} does not describe any individuals making decisions based on police precincts. Another recent book titled \textit{Race Brokers} comes to a similar conclusion that police precincts do not factor into decisions about where one resides \citep{korver2021race}. This is reassurance that police precinct boundaries are not salient. However, police precinct boundaries could still overlap with other, more important government boundaries that do influence where people live. Therefore, we investigate three sub-municipal geographies to see whether they coincide with precincts and whether they affect where people choose to live: counties (aka boroughs), community boards (aka community districts), and school zones (aka catchment areas).}
\begin{figure}[H]
    \centering
    \includegraphics[width=0.5\linewidth]{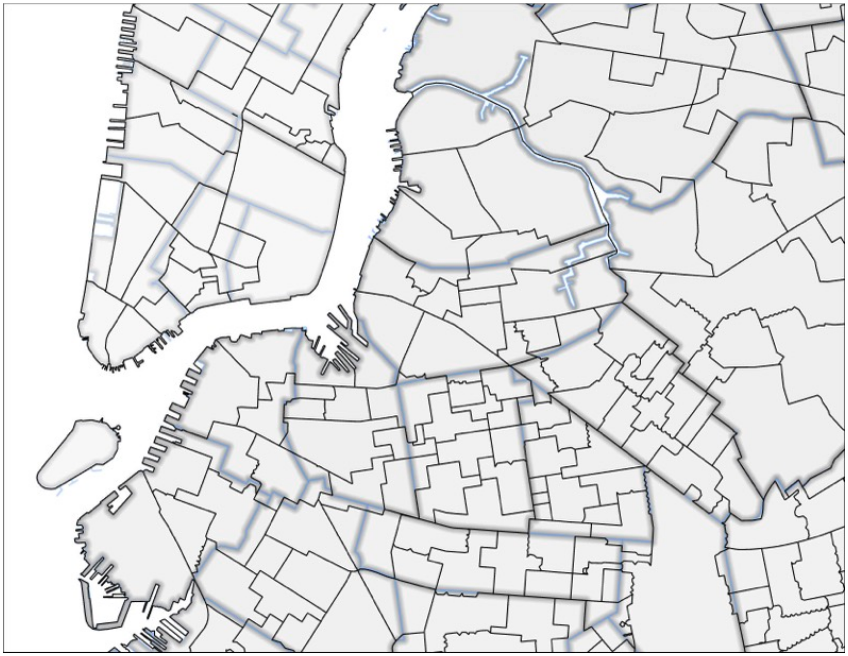}
    \caption{Excerpt of map of NYC police precincts (thick lines with fuzzy gradients) and school zones (solid black lines).}
    \label{fig:precZone}
\end{figure}
\textcolor{black}{
\begin{itemize}
    \item \textbf{Boroughs:} NYC is made up of five counties, also called boroughs: Manhattan, Brooklyn, Queens, the Bronx, and Staten Island. Most of the counties are bordered by water, so those respective boundaries will not appear in any of our analyses. There is one land border, however, between Queens and Brooklyn. Three precincts (75, 83, and 90) fall along that border and share one border with the borough/county. This represents only 2.78\% of all the borders in our data (i.e., 4 out of 144 distinct precinct-precinct borders). Additionally, because NYC has a unified, centralized city government, counties are not responsible for many government functions. The court systems are county-based, but the borough president is a ceremonial role. 
    \item \textbf{Community Districts:} NYC has 59 community boards, local advisory groups. These districts do overlap considerably with police precincts (i.e., many share borders). However, community boards have no final power, only offering recommendations to the city council, mostly on matters of zoning. 
    \item \textbf{School Zones:} While NYC residents are unlikely to consider police precinct, borough, or community district boundaries when choosing where to live, schools play a much larger role in residential decisions \citep{besbris2020upsold}. To investigate the effect of schools, we look at the overlap between police precinct boundaries and school zones. In NYC, middle and high schools are determined by application and residential location plays only a secondary role. Students at these levels often commute quite long distances from home, so we focus on elementary schools here, which are more determined by location. There are 1,048 elementary school zones in the city, each including one elementary school. No elementary school zones are perfectly coterminous with a police precinct, but, as Figure \ref{fig:precZone} shows, some do share borders with police precincts. However, even with school zones, residence is not a guarantee of attendance. Many of the more desirable school zones have more students who want to attend their school than places for them. Spots in these elementary schools are allocated by lottery, and students can also apply to schools outside their zones. If there are open spots in those schools, students living in other zones are allowed to attend. This leads to considerable variation between residence and school attended. 
\end{itemize}
Overall, these findings point to the fact that either police precincts do not share borders with other important boundaries, or these other boundaries are not important enough within NYC to drive someone's decision about where to live. However, despite our reasoning above, we must acknowledge that it is ultimately unverifiable whether precincts influence where individuals decide to live, and a violation of this assumption could affect the validity of our findings.}
}
\end{document}